# Shape- and Cation-Engineered Ferrite Nanoparticles for Enhanced Theranostic Performance in Ovarian Cancer Tumor-Mimicking Phantom

Bahareh Rezaei[1], Md Shahriar[2], Shahriar Mostufa[1], Karla Mercedes Paz González[3], Nguyen Thuy Linh Tran[3], Changxue Xu[2], Jenifer Gómez-Pastora[3], Kai Wu[1,4,*]

[1]Department of Electrical and Computer Engineering, Texas Tech University, Lubbock, TX 79409, USA

[2]Department of Industrial, Manufacturing, and Systems Engineering, Texas Tech University, Lubbock, TX 79409, USA

[3]Department of Chemical Engineering, Texas Tech University, Lubbock, TX 79409, USA

[4]Department of Physics, University of South Florida, Tampa, FL 33620, USA

*Corresponding Author: kaiwu@usf.edu (K.W.)

**Abstract**

Magnetic hyperthermia integrated with magnetic resonance imaging (MRI) requires nanoparticles that combine strong heating, T2 contrast, and hemocompatibility after systemic administration. Here, we prepared citrate-stabilized ferrite nanoparticles ($Fe_3O_4$, $Co_{0.6}Fe_{2.4}O_4$, $Zn_{0.3}Fe_{2.7}O_4$, and $Zn_{0.35}Mn_{0.25}Fe_{2.4}O_4$) in spherical (8-10.5 nm) and quasi-cubic (9-12.5 nm) forms, along with 35 nm $Fe_3O_4$ cubes. By coupling morphology engineering with spinel-lattice cation substitution, we tuned saturation magnetization, coercivity, T2-weighted signal attenuation, and heat dissipation in ovarian tumor-mimicking phantoms under clinically compatible alternating magnetic fields. The 35 nm $Fe_3O_4$ cube and quasi-cubic $Zn_{0.35}Mn_{0.25}Fe_{2.4}O_4$ generated the highest phantom temperature rises, whereas smaller spherical particles produced softer heating. All formulations were cytocompatible in SKOV3 cells, with >96% viability at 24 h. However, hemocompatibility distinguished the leading candidates: 35 nm $Fe_3O_4$ cubes induced red blood cell deformation despite minimal hemolysis, while 9 nm quasi-cubic $Zn_{0.35}Mn_{0.25}Fe_{2.4}O_4$ showed no change versus PBS up to 1,000 µg $mL^{-1}$. Perfusable GelMA models under physiological flow further confirmed biocompatibility. MRI phantoms showed that $Fe_3O_4$ cubes produced extended signal voids that may obscure boundaries. Overall, 9.5 nm quasi-cubic $Zn_{0.35}Mn_{0.25}Fe_{2.4}O_4$ is the leading intravenous candidate, combining strong heating, T2 contrast, and hemocompatibility, whereas large $Fe_3O_4$ cubes are better suited for localized intratumoral hyperthermia.

## 1. Introduction

Precision oncology demands reproducible, quantitative imaging to localize therapeutic targets, deliver focal therapy with accuracy, and verify dose in real time. In the United States, 2,114,850 new cancer cases and 626,140 deaths were projected for 2026, and many patients still face late diagnoses, metastatic progression, treatment resistance, and inequitable access to care [1]. These challenges underscore the need for platforms that integrate diagnosis, therapy delivery, and treatment monitoring into a single quantitative workflow. Among the major gynecologic cancers, ovarian cancer is widely regarded as the most lethal, exhibiting a particularly unfavorable mortality-to-incidence pattern [2-3]. Globally, epithelial ovarian carcinoma (EOC) represents over 90% of cases, with high-grade serous carcinoma as the predominant histotype [2, 4-5]. Despite advances in surgery and systemic chemotherapy, the five-year survival rate for advanced-stage disease remains below 30% [6-7]. Late detection is the primary barrier to cure, and approximately 70% of patients are diagnosed with FIGO (International Federation of Gynecology and Obstetrics) stage III or IV, when metastasis has already occurred within the peritoneal cavity [8-9]. The absence of early clinical symptoms, lack of effective screening biomarkers, and intratumoral heterogeneity contribute to this diagnostic delay. Moreover, the complex vascular and stromal microenvironment of ovarian tumors often impedes drug delivery, fostering treatment resistance and recurrence [10-11].

Conventional diagnostic and therapeutic approaches face substantial limitations in managing such complex diseases. Current imaging modalities, including computed tomography (CT) and positron emission tomography (PET), rely on ionizing radiation, which restricts repeated use for longitudinal monitoring and poses safety concerns for reproductive-age or pregnant patients [12-13]. Magnetic Resonance Imaging (MRI), while radiation-free and capable of high-resolution soft-tissue visualization, relies on gadolinium-based contrast agents that carry nephrotoxicity risks and provide only indirect, non-quantitative information about therapeutic distribution [14]. On the treatment side, systemic chemotherapy and radiotherapy remain systemic and non-selective, often damaging healthy tissues, inducing severe side effects, and limiting cumulative dose [15-16]. Consequently, there is a critical need for image-integrated therapeutic platforms capable of delivering localized, controllable, and biologically safe interventions within deep tumor sites such as the ovary.

Magnetic hyperthermia therapy (MHT) offers a minimally invasive route to achieve localized tumor ablation or sensitization by elevating tumor temperature into the therapeutic window of 42-46 °C [17-18]. When magnetic nanoparticles (MNPs) are exposed to alternating magnetic fields (AMFs), heat is generated through Néel and Brownian relaxation or hysteresis losses, enabling precise spatial confinement of the thermal dose [19]. By incorporating iron-oxide-based MNPs that also act as MRI contrast agents [20], MHT can be integrated with MRI to noninvasively localize and monitor nanoparticle accumulation and retention based on biodistribution. The dual functionality of MNPs as both heat mediators and MRI tracers offers a unique opportunity to merge diagnosis and therapy within a single nanoscale platform [21-22]. However, the therapeutic success of MRI-integrated MHT depends not only on achieving sufficient and controllable heat generation but also on ensuring safe biological interactions, including efficient tumor cell uptake, high cell viability in non-target tissues, and minimal perturbation of blood components. Hemocompatibility, particularly preservation of RBC membrane integrity against hemolysis and morphological deformation, remains a critical consideration since nanoparticle-induced hemolysis or erythrocyte deformation can compromise systemic safety during intravenous administration [23]. Moreover, the morphology, size, composition, and surface chemistry of MNPs strongly dictate both their magnetic performance and their biological fate, often presenting a trade-off between heating efficiency and biocompatibility [24].

To address these challenges, the present study engineers shape- and composition-tuned ferrite nanoparticles and systematically evaluates their physicochemical, magnetic, and biological profiles. By comparing spherical and quasi-cubic morphologies across $Fe_3O_4$, $Co_{0.6}Fe_{2.4}O_4$, $Zn_{0.3}Fe_{2.7}O_4$, and $Zn_{0.35}Mn_{0.25}Fe_{2.4}O_4$ systems, we evaluate how anisotropy and cation substitution influence heating behavior under AMFs, MRI contrast, cellular uptake, viability, and RBC morphology. To further emulate the vascularized tumor microenvironment and evaluate nanoparticle biocompatibility under physiological flow, we also employed a 3D perfused GelMA-gelatin construct as a vascular analog. This system enables continuous medium perfusion and oxygen diffusion, supporting evaluation of MNP-cell interactions under perfusion-driven gradients that approximate key aspects of *in vivo* exposure following intravenous administration [25]. By coupling quantitative MRI contrast measurements, controlled nanoparticle heating, and rigorous biocompatibility assessment, this work establishes a preclinical framework for MRI-integrated magnetic hyperthermia in ovarian tumor-mimicking phantom models.

## 2. Results and Discussions

### 2.1. Morphological Control and Structural Characterization of Quasi-Cubic and Spherical Ferrite Nanoparticles

To achieve precise morphological control across the ferrite compositions, two complementary aqueous routes, co-precipitation and hydrothermal synthesis, were developed and optimized. The co-precipitation method (**Fig. 1A**), designed to produce quasi-cubic ferrite nanoparticles, was conducted by simultaneously precipitating $Fe^{3+}$, $Fe^{2+}$, and dopant cations ($Co^{2+}$, $Zn^{2+}$, $Mn^{2+}$) in an alkaline environment using NaOH as the precipitating agent. Rapid nucleation under high supersaturation is consistent with a burst-nucleation scenario, in which nucleation occurs over a short time window followed by comparatively slower crystal growth [26-30]. Citric acid was introduced after NaOH primarily to adsorb onto the as-formed nanoparticle surfaces and provide electrostatic stabilization, thereby controlling post-synthesis aggregation rather than strongly directing the initial faceting [31-36]. The resulting particles exhibited quasi-cubic shape with average core sizes of 12.5±1.8 nm for C1, 10±2 nm for C2, 9.5±1.5 nm for C3, and 10.5±1.8 nm for C4, confirming controlled anisotropic growth across the ferrite series (**Fig. 1B**). For clarity and consistency across the characterization and biological assays, the 9 ferrite nanoparticle samples are referenced by fixed IDs (C1-C4, S1-S4, and C35). **Table 1** summarizes the MNP composition, morphology, and core size metrics used for all subsequent comparisons.

**Table 1.** Composition, synthesis route, morphology, and core size of the ferrite nanoparticle library.

| Sample Stoichiometry | Method of Synthesis | Shape | Core Size | IDs |
|---|---|---|---|---|
| $Fe_3O_4$ | Hydrothermal | Spherical | 8.8±1.3 | S1 |
| $Zn_{0.3}Fe_{2.7}O_4$ | Hydrothermal | Spherical | 8±1.2 | S2 |
| $Zn_{0.35}Mn_{0.25}Fe_{2.4}O_4$ | Hydrothermal | Spherical | 8.5±1.8 | S3 |
| $Co_{0.6}Fe_{2.4}O_4$ | Hydrothermal | Spherical | 10.5±1.1 | S4 |
| $Fe_3O_4$ | Coprecipitation | Quasi-Cubic | 12.5±1.8 | C1 |
| $Zn_{0.3}Fe_{2.7}O_4$ | Coprecipitation | Quasi-Cubic | 10±2 | C2 |
| $Zn_{0.35}Mn_{0.25}Fe_{2.4}O_4$ | Coprecipitation | Quasi-Cubic | 9.5±1.5 | C3 |
| $Co_{0.6}Fe_{2.4}O_4$ | Coprecipitation | Quasi-Cubic | 10.5±1.8 | C4 |
| $Fe_3O_4$ | Hydrothermal | Cubic | 35±3.2 | C35 |

In contrast, the hydrothermal method produced highly uniform, spherical ferrite nanoparticles across all compositions when reactions were carried out in sealed Teflon-lined autoclaves (**Fig. 1C**). Using $NH_4OH$, the solution pH was tuned within a narrow window, which proved critical for consistent hydrolysis and incorporation of divalent ($Co^{2+}$, $Zn^{2+}$, $Mn^{2+}$) and trivalent ($Fe^{3+}$) ions, as reflected by the spinel XRD patterns [37-38] (**Fig. 1F**) and narrow TEM size distributions (**Fig. 1D**). Elevated temperature and pressure promote rapid hydrolysis and uniform mass transport, which can favor isotropic growth [39], with average diameters of 8.8±1.3 nm for S1, 8±1.2 nm for S2, 8.5±1.8 nm for S3, and 10.5±1.1 nm for S4 (**Fig. 1D**) [40-42]. Citric acid acts as an in-situ capping agent stabilized the colloids electrostatically, preventing agglomeration without introducing shape anisotropy[30, 43].

On the other hand, $Fe_3O_4$ nanoparticles synthesized hydrothermally at 195 °C for 16 h exhibited a more faceted, nearly cubic morphology, with an average size of 35±3.2 nm. This behavior may be attributed to a shift toward more thermodynamic control at elevated temperature and prolonged reaction time, where growth and Ostwald ripening become increasingly influential, and facet stabilization by solution species can shift the relative stability and growth rates of low-index surfaces (e.g., (111) and (001)), enabling well-defined cubic morphologies [44-46]. Notably, the stability of magnetite low-index facets depends on surface termination and the hydrothermal chemical environment; theoretical studies commonly identify the {100} and {111} families as among the most stable, consistent with cubic equilibrium morphologies. Because the composition is undoped, dopant-driven strain and cation redistribution effects are minimized, allowing the nanocrystals to evolve toward a more thermodynamically favored faceted shape as differences between the {100}, {110}, and {111} surface free energies manifest during prolonged hydrothermal growth [44, 47-48].

TEM images reveal that the hydrothermal method-prepared MNPs (**Fig.1D**) are predominantly spherical with narrow size distributions, whereas the co-precipitation method-prepared MNPs (**Fig. 1B**) exhibit clear faceting and quasi-cubic geometry. Particle-size histograms confirm mean diameters of around 9.5-12.5 nm for the quasi-cubic samples and 8-10.5 nm for the spherical ones. The hydrothermally synthesized, undoped C35 sample distinctly displays a uniform cubic morphology with an average size of 35 nm, consistent with the thermodynamic facet-stabilization mechanism.

XRD patterns (**Fig. 1E&F**) confirmed the single-phase spinel structure for all samples. Relative to undoped $Fe_3O_4$, the small but systematic shifts in peak positions reflect composition-dependent changes in the lattice parameter (a) **(Table 2)**, which primarily arise from substitution-induced size mismatch and possible cation redistribution and stoichiometry effects. Doping of larger divalent cations such as $Mn^{2+}$ (ionic radius of 0.83 Å) generally expands the lattice and shifts reflections toward lower 2θ values [49-50]. Zn-containing ferrites can also exhibit lattice expansion because the ionic radius of $Zn^{2+}$ is larger than that of $Fe^{3+}$, and its site preference can modify the average cation-oxygen bond lengths and unit-cell parameter through charge compensation and redistribution of $Fe^{2+}/Fe^{3+}$ between A and B sites. In contrast, Co-substituted ferrites often show a smaller net shift relative to $Fe_3O_4$ because $Co^{2+}$ typically prefers B sites in inverse-spinel cobalt ferrite and, in magnetite-based spinel, $Co^{2+}$ can effectively replace octahedral $Fe^{2+}$; since $Co^{2+}$ ionic radius (0.745 Å) is relatively close to $Fe^{2+}$ ionic radius (0.78 Å), the resulting change in unit-cell size can be modest compared with $Mn^{2+}$ substitution [51-52].

**Table 2.** Lattice parameters of the samples.

| IDs | Lattice parameter (Å) | IDs | Lattice parameter (Å) |
|---|---|---|---|
| **S1** | 8.3317 | **C1** | 8.3320 |
| **S2** | 8.3741 | **C2** | 8.3859 |
| **S3** | 8.3865 | **C3** | 8.4021 |
| **S4** | 8.3432 | **C4** | 8.3667 |
| - | - | **C35** | 8.3651 |

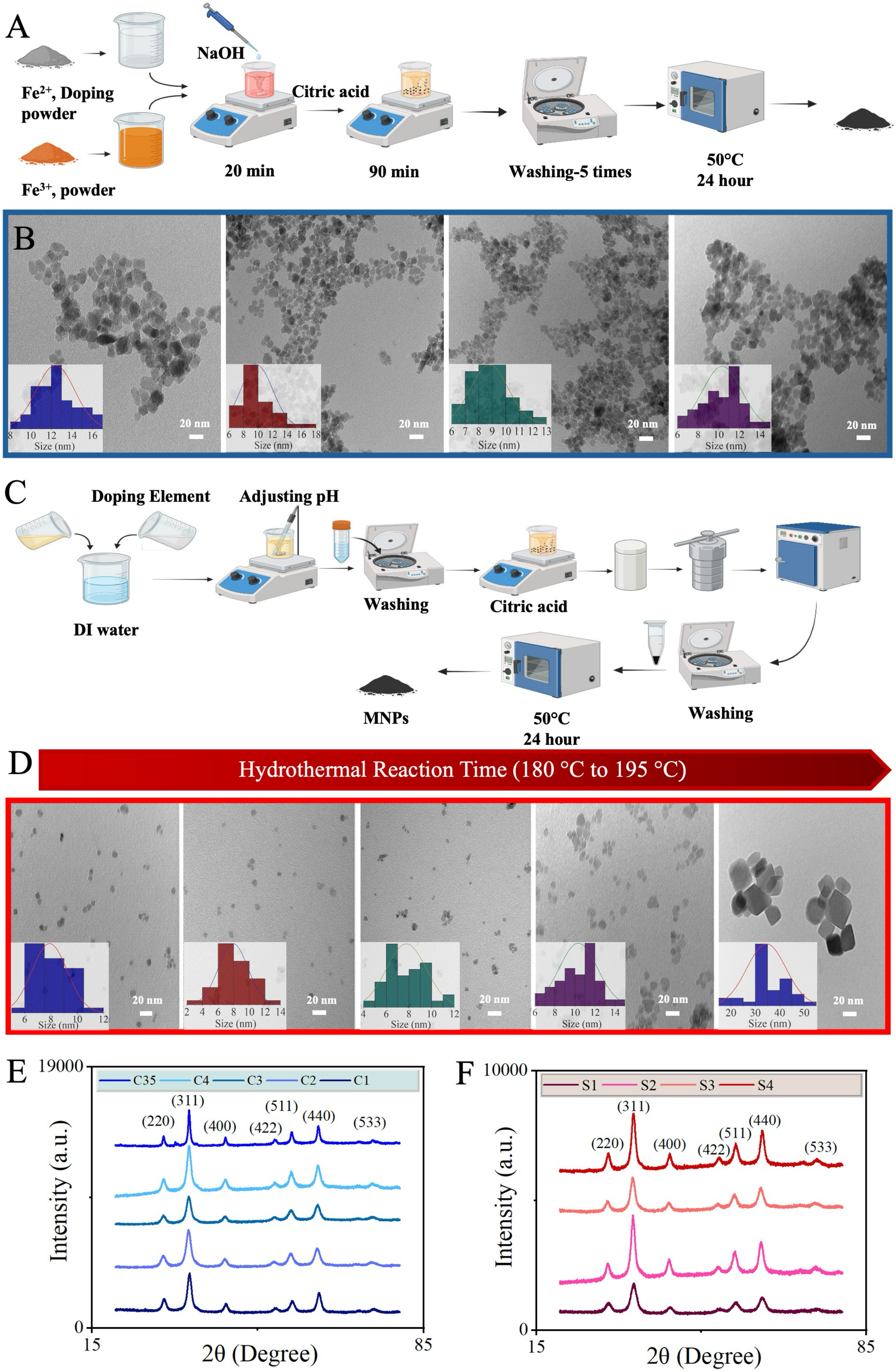
A
Fe²⁺, Doping powder
Fe³⁺, powder
NaOH
Citric acid
20 min
90 min
Washing-5 times
50°C
24 hour
B
20 nm
Size (nm)
C
Doping Element
Adjusting pH
DI water
Washing
Citric acid
Washing
50°C
24 hour
MNPs
D
Hydrothermal Reaction Time (180 °C to 195 °C)
E
Intensity (a.u.)
2θ (Degree)
C35
C4
C3
C2
C1
(220)
(311)
(400)
(422)
(511)
(440)
(533)
F
S1
S2
S3
S4
Intensity (a.u.)
2θ (Degree)

**Fig. 1.** Dual-route synthesis and structural characterization of the ferrite nanoparticle library. (A) Schematic of the co-precipitation workflow used to synthesize quasi-cubic ferrite MNPs. (B) representative TEM images and corresponding size distributions for the quasi-cubic series (C1-C4). Scale bar: 20 nm. (C) Schematic of the hydrothermal workflow used to synthesize spherical ferrite MNPs. (D) Representative TEM images and size distributions for the spherical series (S1-S4) and the large C35 obtained under extended hydrothermal reaction conditions (arrow indicates increasing hydrothermal reaction temperature window). Scale bar: 20 nm. (E) XRD patterns confirming the spinel phase for the quasi-cubic series (C1-C4 and C35) and (F) the spherical series (S1-S4).

### 2.2. Physicochemical Characterization and Colloidal Stability of Ferrite Nanoparticles

Dynamic light scattering (DLS) and ζ-potential measurements were performed in deionized water to evaluate the colloidal behavior and electrostatic stability of the synthesized ferrite nanoparticles (**Fig. 2A&B&E**). As expected, the hydrodynamic diameters ($D_h$) obtained from DLS were larger than the core sizes measured by TEM, reflecting the presence of the citrate layer on the nanoparticle surfaces and dynamic interparticle interactions in suspension [20, 53]. Spherical nanoparticles exhibited smaller hydrodynamic diameters (36-42 nm), whereas quasi-cubic nanoparticles showed larger $D_h$ values (43-76 nm). Given the anisotropic morphology of the quasi-cubic samples, these differences may reflect a combination of shape effects, stronger interparticle interactions, and a greater tendency toward transient association in water. The C35 and C4 samples displayed the largest hydrodynamic size, consistent with partial clustering in aqueous suspension.[54-57]

ζ-potential measurements confirmed negatively charged particle surfaces across all ferrite formulations, with values ranging from -16 to -61.3 mV (**Fig. 2E**). FTIR spectra confirmed citrate functionalization on the nanoparticle surface through characteristic citrate-related carboxylate vibrations, including asymmetric and symmetric $COO^-$ stretching bands near 1585 and 1380 $cm^{-1}$, respectively, together with additional citrate-associated features in the fingerprint region near 960 and 800 $cm^{-1}$ (**Fig. 2C&D**). EDX analysis additionally revealed a detectable carbon signal, supporting the presence of an organic surface layer and providing corroborative evidence for citrate coating alongside FTIR and ζ-potential (**Fig. 2E&H&I; supplementary materials Fig. S1 & Fig. S2**) [58-59]. Among the quasi-cubic samples, C3 exhibited the most negative ζ-potential (-61.3 mV), consistent with strong electrostatic stabilization. The other quasi-cubic nanoparticles (C1, C2, and C4) also exhibited strongly negative ζ-potentials of -43.7, -48.3, and -44.6 mV, respectively, which are generally sufficient to maintain stable dispersions in deionized water. The spherical nanoparticles (S1 to S4) showed moderately negative ζ-potentials ranging from -27.3 to -38.9 mV, consistent with stable aqueous dispersions and their comparatively smaller hydrodynamic diameters. In contrast, the large cubic $Fe_3O_4$ sample (C35) displayed a substantially reduced ζ-potential magnitude (-16 mV), a range commonly associated with only short-term colloidal stability and increased aggregation propensity, which plausibly contributes to its larger $D_h$ in water. Using the conventional guideline that $|\zeta|$ values above about 30 mV often indicate good electrostatic stabilization, while $|\zeta|$ values below 20 mV are typically associated with limited short-term stability, the ζ-potential data agree well with the 12 h sedimentation observations (**Fig. 2G**), in which highly charged formulations remained dispersed while C35 samples showed more visible settling [60]. Overall, these results indicate that colloidal behavior in deionized water is governed by the combined effects of surface charge, particle morphology, and magnetic dipolar interactions.

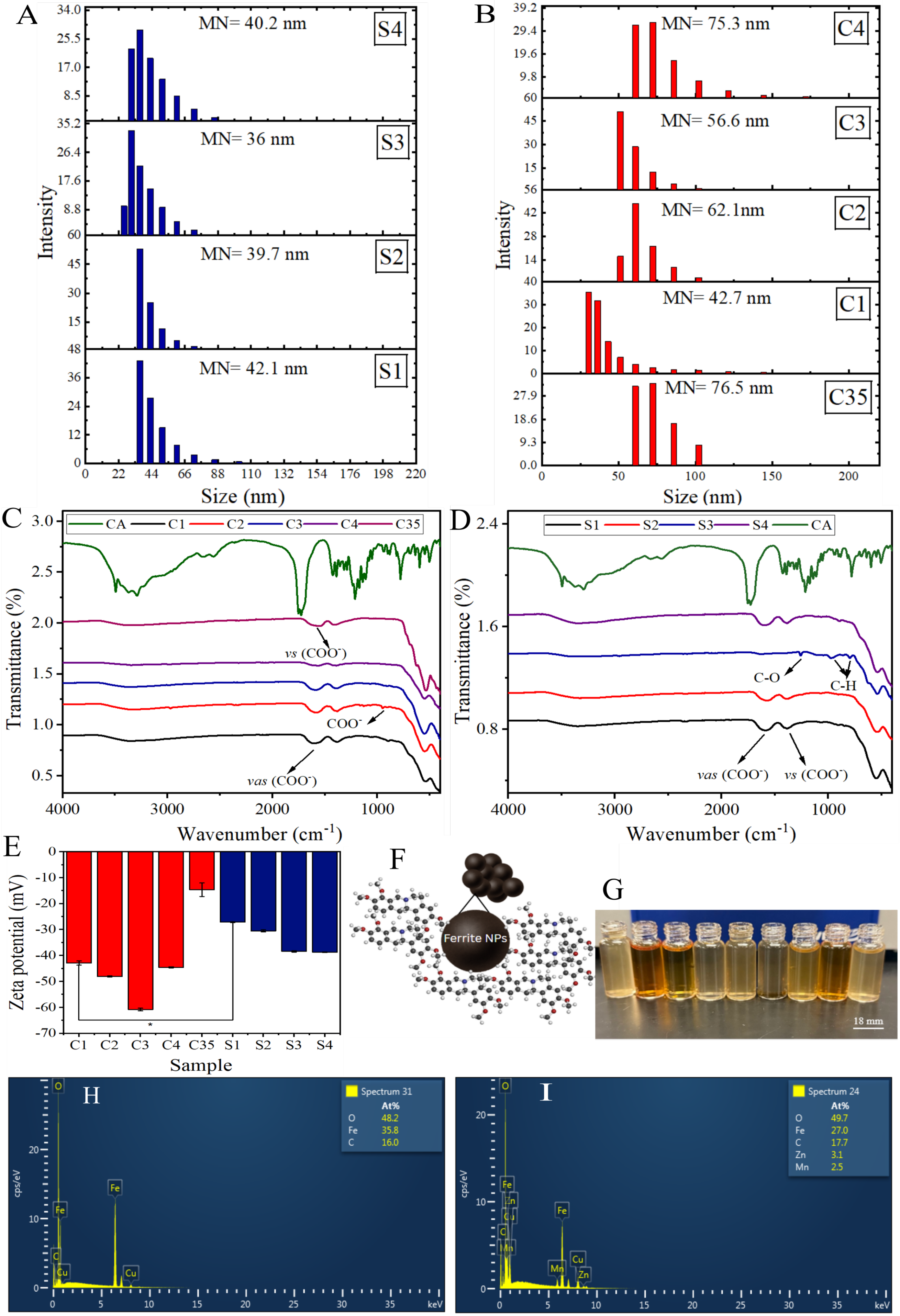

A
MN= 40.2 nm
S4
MN= 36 nm
S3
MN= 39.7 nm
S2
MN= 42.1 nm
S1
Intensity
Size (nm)
B
MN= 75.3 nm
C4
MN= 56.6 nm
C3
MN= 62.1nm
C2
MN= 42.7 nm
C1
MN= 76.5 nm
C35
Intensity
Size (nm)
C
CA C1 C2 C3 C4 C35
vs (COO-)
COO-
vas (COO-)
Transmittance (%)
Wavenumber (cm-1)
D
S1 S2 S3 S4 CA
C-O
C-H
vas (COO-)
vs (COO-)
Transmittance (%)
Wavenumber (cm-1)
E
Zeta potential (mV)
C1 C2 C3 C4 C35 S1 S2 S3 S4
Sample
F
Ferrite NPs
G
18 mm
H
Spectrum 31
At%
O 48.2
Fe 35.8
C 16.0
cps/eV
keV
I
Spectrum 24
At%
O 49.7
Fe 27.0
C 17.7
Zn 3.1
Mn 2.5
cps/eV
keV

**Fig. 2.** Hydrodynamic size, surface chemistry, and colloidal stability of citrate-coated ferrite MNPs. DLS size distributions of spherical (A) and quasi-cubic (B) nanoparticle samples. FTIR spectrum of the spherical (C) and quasi-cubic (D) nanoparticles. (E) Zeta potential values of the samples, with statistically significant differences between the indicated selected groups ($p < 0.05$). (F) Schematic representation of citrate-coated MNPs. (G) Photograph of all samples after 12 h, indicating suspension colloidal stability. EDX spectrum of sample C35 (H) and C3 (I). MN represents the number-weighted mean diameter shown in each panel.

**2.3. Cytocompatibility and Intracellular Dynamics of Ferrite Nanoparticles in SKOV3 Cells**

The interaction between the engineered ferrite nanoparticles and cellular membranes was investigated through comprehensive cytocompatibility and uptake studies in SKOV3 ovarian cancer cells. Live/Dead fluorescence imaging enabled quantitative assessment of cell viability via image-based counting of live cells relative to the total number of cells. SKOV3 cultures were incubated with nanoparticle dispersions at concentrations ranging from 100 to 500 µg $mL^{-1}$ for 24 h under standard conditions. **Fig. 3A** schematically illustrates the experimental workflow, outlining nanoparticle exposure, staining, and microscope-based imaging used to assess viability and cellular uptake. **Fig. 3B** summarizes cell viability across all formulations at 100 and 500 µg $mL^{-1}$, and **Fig. 3C** presents representative cellular uptake images acquired at 100 µg $mL^{-1}$. Representative fluorescence images show that most SKOV3 cells retained normal epithelial-like morphology and green fluorescence after nanoparticle exposure, consistent with high viability (>96–99%) at 24 h across all ferrite compositions. The absence of red nuclei indicates preserved membrane integrity and minimal acute membrane-compromising cell death after nanoparticle exposure [61-63]. This observation is consistent with prior reports [63-64] that citrate-stabilized, negatively charged nanoparticle surfaces tend to exhibit reduced nonspecific electrostatic interactions with cell membranes and lower acute cytotoxicity relative to positively charged analogues. In our system, citrate stabilization yielded negative surface potentials ($\zeta$= -16 to -61.3 mV), which may contribute to reduced membrane disruption and preserved viability. A marginal increase in dead cells was observed for the S4 and C4 samples at 500 µg $mL^{-1}$ concentration, resulting in a slight reduction in overall viability (2.5-3.5%) [65-66]. Nevertheless, most cells remained adherent with intact nuclei and preserved morphology at the 24 h timepoint [67-71].

Microscopic examination after a 48-h incubation revealed composition-, size-, and morphology- dependent differences in cell-associated accumulation (**Fig. 3C**). Among all samples, C35 exhibited the highest highest cell-associated accumulation after 48 h, consistent with size-shape effects on membrane wrapping and with theoretical models predicting an optimal nanoparticle size window of 25-35 nm for relatively rapid endocytic internalization [72-75]. Among the small quasi-cubic samples, C3 and C4 exhibited higher cellular uptake than C1. Although these formulations are closely matched in morphology and near-matched in core size, their uptake did not track simply with electrostatic charge: C3 showed the most negative $\zeta$-potential, yet C4 displayed even higher uptake. This lack of correlation suggests that additional composition-dependent interfacial processes in serum-containing media, including differences in protein corona formation and colloidal behavior, may contribute to the observed differences in cellular association. Such changes can alter the biological identity of the nanoparticles and their interactions with cell-surface receptors, independently of the initial surface charge [76-79]. The lower uptake of C2 relative to C3, despite similar morphology and comparable $\zeta$-potential, further supports this interpretation that the absence of $Mn^{2+}$ surface sites in C2 may result in a distinct protein corona composition that is less favorable for cell association. C4's higher uptake relative to C3, despite lower surface charge, may likewise reflect differences in its biological identity and serum-protein interactions arising from ferrite composition [80-83]. Among the spherical samples, S4 and S3 showed the highest uptake, consistent with a possible contribution from composition-dependent protein corona formation mirroring the trends observed from quasi-cubic counterparts[82]. Collectively, these trends suggest that uptake within both the spherical and quasi-cubic samples is governed by coupled effects of ferrite composition and interfacial and dispersion behavior in serum-containing media, rather than size, ζ-potential, or morphology alone.

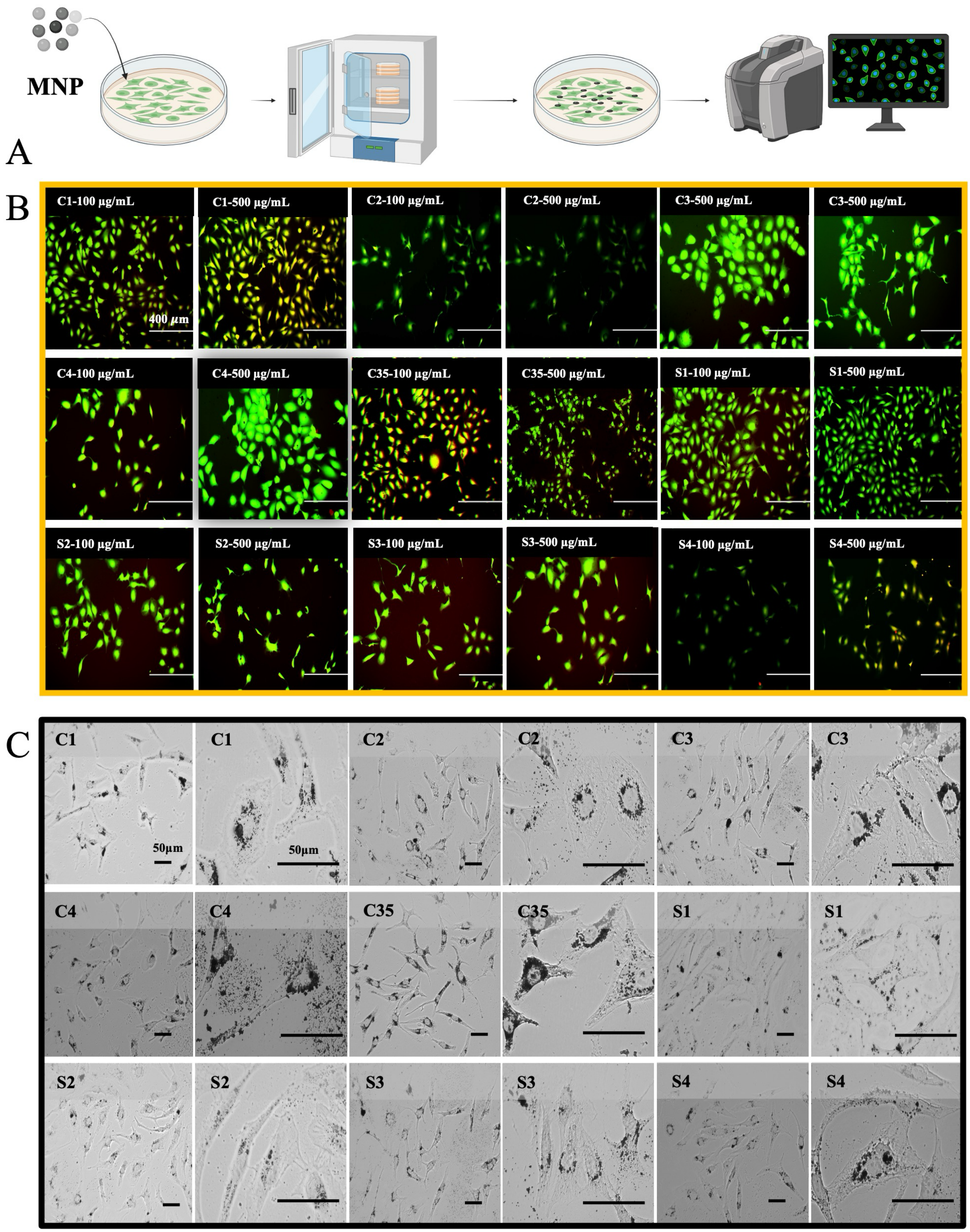

**Fig. 3.** Live/Dead assay and cellular internalization of ferrite MNPs in SKOV3 cells. (A) Schematic illustration of the experimental workflow in Live/Dead viability assessment and cellular uptake studies. (B) Live/Dead assay results of SKOV3 cells after 24 h incubation with the nanoparticle samples at concentrations of 100 and 500 µg

mL$^{-1}$. Scale bar: 400 μm. (C) Cellular uptake of the nanoparticles in SKOV3 cells after 24 h incubation at 100 μg mL$^{-1}$. Scale bar: 50 μm. Schematics created in BioRender.com.

**2.4. Magnetic Hyperthermia Performance in Biological and Tumor-Mimicking Phantom**

Having established the biocompatibility and cellular uptake profiles of the synthesized ferrite nanoparticles, the next objective was to determine which formulations combine strong magnetic heating performance with biological safety for MRI- integrated MHT. Accordingly, the static magnetization properties of all the samples were first quantified using a Physical Property Measurement System (PPMS), providing saturation magnetization ($M_s$) and coercivity ($H_c$) as baseline descriptors of magnetic responses. Static magnetic hysteresis (M-H) loops were measured over a field range of ±1.5 T. **Fig. 4A** summarizes the M-H loops of the quasi-cubic nanoparticles (C1-C4) and 35 nm $Fe_3O_4$ cubic nanoparticle (C35), while **Fig. 4C** summarizes those of the spherical samples (S1-S4). The zoomed-in M-H loops (**Fig. 4B&D**) show that, at room temperature, all samples except C35, C4, and S4 exhibited low coercivity, consistent with predominantly superparamagnetic behavior.

The magnetic hyperthermia performance of all samples was then evaluated at a fixed nanoparticle concentration of 7 mg mL$^{-1}$ to enable direct mechanistic comparison of medium-dependent heating behavior under identical concentration conditions. This concentration was selected to provide a biologically relevant and experimentally practical nanoparticle loading consistent with concentrations previously used in preclinical and tumor-relevant magnetic hyperthermia studie[84-85 86-87]. The experimental setup used for all magnetic hyperthermia measurements is schematically illustrated in **Fig. 5A**. Experiments were performed in fetal bovine serum (FBS)-supplemented SKOV3 medium containing cells and in SKOV3-embedded agar phantoms under an AMF of 30 mT and 101 kHz, a regime used in preclinical magnetic hyperthermia and within extended safety windows for localized exposure ($H \cdot f \leq 5 \times 10^9$ $A \cdot m^{-1} \cdot s^{-1}$)[88]. In FBS media, heating showed a clear dependence on particle morphology and magnetic properties (**Fig. 4F**). The observed trends are consistent with smaller particles heating mainly through Néel-relaxation losses under constrained conditions, while larger cubes may additionally benefit from hysteresis-related losses depending on their anisotropy, magnetic volume, and interaction state [57, 89-90]. The strong performance of C35 is consistent with extensive reports that iron-oxide nanocubes can achieve high heating efficiencies under clinically relevant fields due to their magnetic volume, enhanced shape anisotropy, and shape anisotropy-driven dynamic losses [91-92]. Importantly, the high heating performance of C3, despite its low coercivity, is also expected because superparamagnetic nanoparticles can still dissipate substantial power under an AMF when their effective Néel relaxation dynamics and interaction state are favorably matched to the field frequency range, particularly in viscous or biologically constrained environments where Brownian rotation is reduced [93-100]. In contrast, cobalt ferrite, represented by C4 and S4, is a high-anisotropy system; its high magnetocrystalline anisotropy may shift the Néel relaxation time outside the optimal range at 101 kHz, while the field amplitude of 30 mT may not be sufficient to efficiently drive hysteresis losses in Co-ferrite. Consequently, both experiments and modeling show that the absorbed power of Co-ferrites depends sensitively on viscosity, immobilization, interparticle interactions, and whether the applied field amplitude can generate a sufficiently large dynamic loop area, confirming that magnetically harder nanoparticles do not necessarily produce higher heating at a fixed field amplitude of 30 mT [95, 101]. The lower temperature rises of the spherical samples relative to quasi-cubic or cubic counterparts are therefore consistent with their smaller magnetic core volumes and lower shape anisotropies, which typically yield smaller dynamic losses at a fixed field even when $M_s$ values are comparable [102]. For the Zn-ferrites, the weaker heating compared to Mn-ferrites is also consistent with composition-dependent cation distribution effects that alter the dynamic loss efficiency under the AMF; thus, heating does not necessarily scale monotonically with $M_s$ [77, 103-104].

When nanoparticles are embedded in SKOV3 tumor-mimicking agar phantoms, the Brownian motion is strongly suppressed by the gel matrix, and magnetic heating is expected to be governed predominantly by Néel relaxation and hysteresis losses, with reduced contribution from Brownian rotation [105-106]. As shown in **Fig. 4E**, under identical AMF, C35 again yielded the largest temperature rises, followed by C3 and C4, whereas S1, S2, C1, and C2 samples stabilized below the desired hyperthermia temperature, consistent with their weaker heating response in the gel environment.

Consistently, the SAR (specific absorption rate) and ILP (intrinsic loss power) values generally decreased for all formulations upon agar immobilization. The most pronounced reduction was observed for C35. Although this formulation is magnetically blocked, its relatively large hydrodynamic diameter may influence its rotational dynamics and overall magnetic heating response in liquid suspension. These rotational contributions are strongly restricted upon agar immobilization. In addition, the relatively low ζ-potential of C35, approximately -16 mV, may indicate weaker electrostatic stabilization and stronger interparticle interactions in suspension, which can further modify the dynamic magnetic losses.  Once immobilized within the gel matrix, both Brownian rotation and cluster dynamics are restricted, leading to a reduction in SAR (**Fig. 4 G&H**). In contrast, smaller nanoparticles retained a larger fraction of their heating efficiency in agar, consistent with a more Néel-dominated response under immobilized conditions [107]. The sustained advantage of C3 over C4 under immobilization reflects C3's lower effective anisotropy, placing its Néel relaxation time closer to the optimal window under this field frequency [108-110]. Notably, the reduced $\Delta T$ in agar compared with FBS-supplemented SKOV3 medium is commonly observed when nanoparticles are embedded in gel-like media and likely arises from the combined effects of hindered rotational motion, immobilized clustering, and enhanced heat conduction to the phantom [107]. Collectively, these results indicate that both morphology and cation composition modulate heating performance under biologically relevant, rotationally constrained conditions.

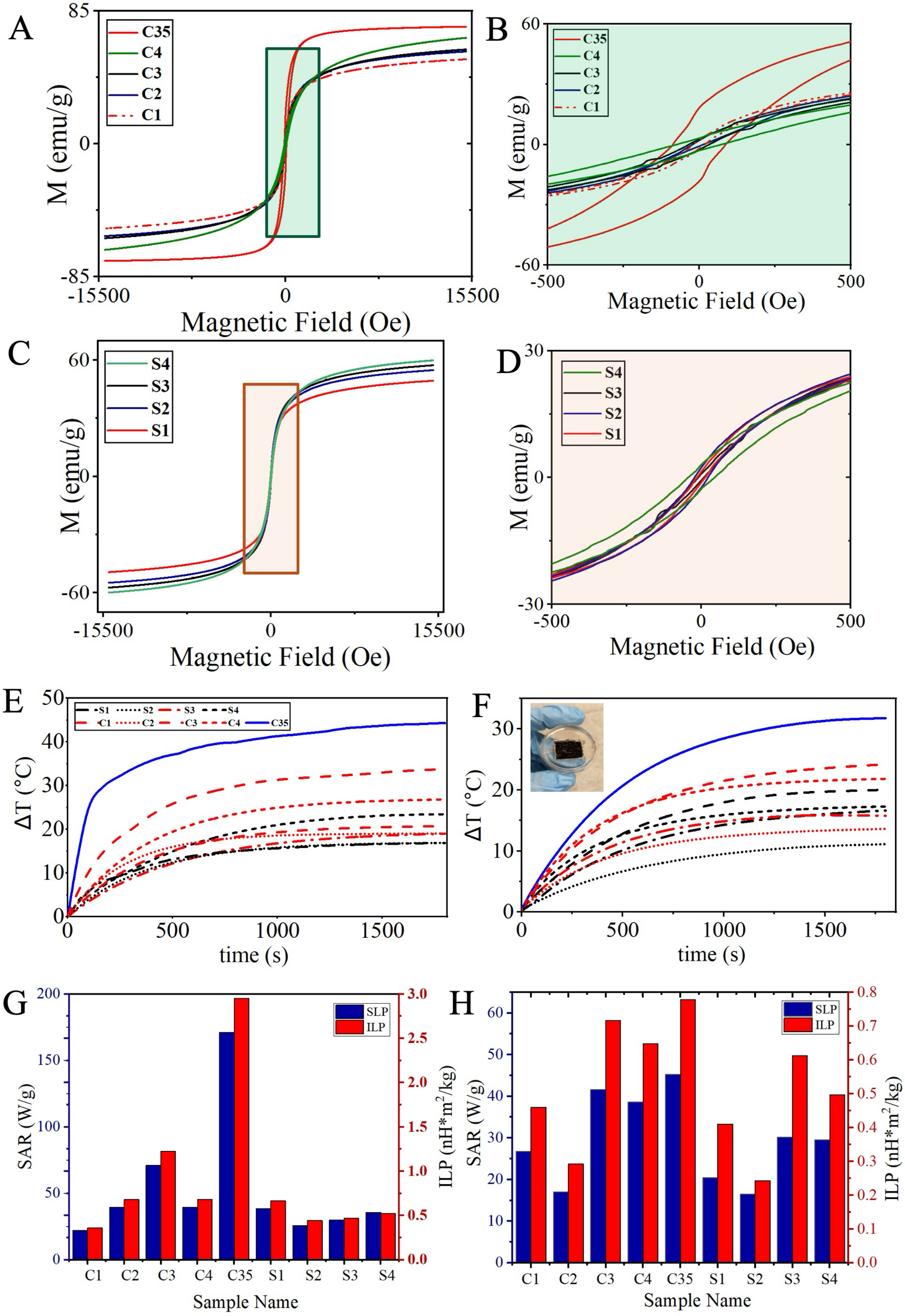
A
B
C
D
E
F
G
H
C35
C4
C3
C2
C1
S4
S3
S2
S1
M (emu/g)
Magnetic Field (Oe)
ΔT (°C)
time (s)
SLP
ILP
SAR (W/g)
ILP (nH*m²/kg)
Sample Name

**Fig. 4.** Magnetic properties and AMF-induced heating behavior of ferrite MNPs with quasi-cubic and spherical morphologies. Static M-H loops of (A) quasi-cubic samples (C1, C2, C3, C4, and C35) and (C) spherical samples (S1, S2, S3, and S4). Zoomed-in view of static M-H loops at a smaller field range of 500 Oe for (B) quasi-cubic samples and (D) spherical samples. Heating profiles of all samples in (E) FBS-supplemented SKOV3 medium and (F) SKOV3 tumor-mimicking agar phantoms. Statistics of SLP and ILP values for all samples measured in (G) FBS-supplemented SKOV3 medium and (H) SKOV3 tumor-mimicking agar phantoms.

After magnetic hyperthermia, cell viability was assessed again using the Live/Dead assay with an untreated control group included for comparison **(Fig. 5B)**. Across formulations, post-AMF viability broadly tracked the delivered thermal dose, with the strongest heaters producing the most pronounced increase in red-stained cells. Notably, viability did not map perfectly onto bulk $\Delta T$ (**Fig. 5C**). For example, S4 produced moderate heating yet retained a higher fraction of green cells. This is expected because bulk temperature is an average metric, whereas cytotoxicity depends on the local temperature-time history at the cell layer and on formulation-dependent nanoparticle-cell association in serum-containing media [111-113]. Beyond heat deposition, spinel ferrites are reported to possess intrinsic enzyme-mimetic redox activity. Under mildly acidic conditions, magnetite-type ferrites can catalyze Fenton/Fenton-like reactions, converting endogenous $H_2O_2$ into highly cytotoxic •OH [114-115]. The magnitude of this peroxidase-like activity depends on both nanoparticles' intrinsic parameters and ROS-relevant microenvironmental factors that govern iron redox cycling and substrate availability. Accordingly, cation engineering can further modulate redox-active surface sites and catalytic efficiency, providing a plausible route by which composition influences ROS-mediated cytotoxicity in addition to magnetic heating [114].

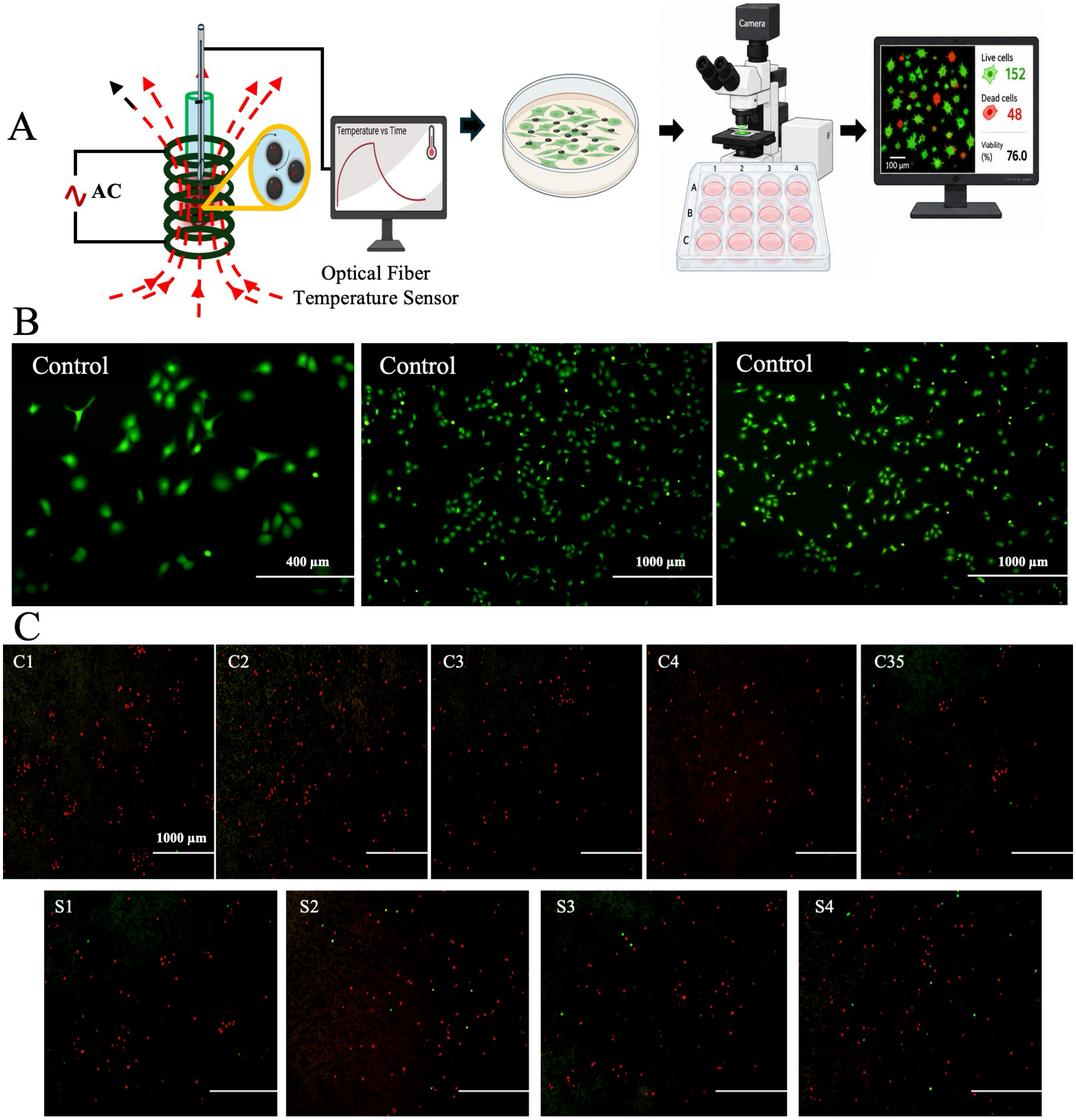

**Fig.5** *In vitro* response of ferrite MNPs with quasi-cubic and spherical morphologies. (A) Schematic representation of the magnetic hyperthermia setup used for AMF exposure and temperature monitoring, followed by Live/Dead viability assessment. (B) Control groups without MNP treatment. (C) Representative Live/Dead fluorescence images of cells following treatment with quasi-cubic and spherical nanoparticle formulations, with live cells shown in green and dead cells in red. Schematics created in BioRender.com.

**2.5. Hemocompatibility and Morphological Response of Red Blood Cells**

To evaluate blood compatibility under clinically relevant systemic exposures, human red blood cells (RBCs) were incubated with C35 and C3, C4, S3 and S4 nanoparticle suspensions at 200 µg mL$^{-1}$ and 1000 µg mL$^{-1}$ for 1 h and 24 h at 37 °C, respectively (**Fig. 6A**). These five formulations were prioritized for hemocompatibility assessment because they generated therapeutically relevant temperature rises in tumor-mimicking agar phantoms under the applied AMF conditions, establishing them as the only candidates within the library capable of delivering meaningful thermal doses in a biologically constrained environment. The lower concentration (200 µg $mL^{-1}$) approximates peak blood levels achieved after intravenous administration of clinically used iron-oxide formulations such as ferumoxytol, where a standard 510 mg Fe dose (≈ 7 mg Fe $kg^{-1}$ in a 70 kg adult) produces a $C_{max}$ of around 200 μg Fe $mL^{-1}$. The higher concentration (1000 μg $mL^{-1}$) was selected to reflect peak blood levels attainable in preclinical theranostic and hyperthermia studies, where systemic or repeated injections of 20-80 mg Fe $kg^{-1}$ iron-oxide nanomaterials are used to achieve sufficient intratumoral accumulation and heating while remaining within tolerated dose ranges in rodents[116]. Thus, the 200-1000 μg $mL^{-1}$ window spans realistic intravenous (IV) exposure levels of nanoparticles in MRI and emerging systemic hyperthermia protocols, providing a conservative basis for hemocompatibility assessment.

Under these concentrations, hemolytic activity, RBC counts, and microscopic morphology were analyzed to determine how particle composition, size, and shape affect membrane integrity and aggregation behavior. All samples exhibited hemolysis levels below the non-hemolytic/slightly hemolytic threshold defined by ASTM F756, classifying them as acceptable from a purely lytic standpoint. As shown in **Fig. 6D,** the C35 sample induced visible distortion of erythrocyte shape, manifesting as *echinocyte* and *stomatocyte* forms after both 1 h and 24 h incubation, particularly at 1000 μg $mL^{-1}$. Although overall hemolysis remained minimal, this deformation indicates mechanical and structural stress rather than membrane rupture [23, 117-118]. Several factors may contribute to this behavior, including mechanical stress caused by sharp cubic edges that exert localized shear and bending forces on the compliant erythrocyte membrane, disturbing curvature equilibrium, and triggering cytoskeletal contraction leading to echinocyte formation[117, 119]. Second, high magnetic moments and anisotropies of C35 nanoparticles promote spontaneous dipolar coupling even in the absence of an applied field. Even under static conditions, small magnetic clusters can form via magnetostatic attraction, transiently trapping RBCs and applying localized compressive stress that amplifies shape alteration [120]. Third, the relatively low ζ-potential of C35 (-16 mV) reflects insufficient surface citrate coverage, reducing electrostatic repulsion against the negatively charged erythrocyte membrane and allowing closer particle-membrane interactions, which amplifies the mechanical stress imposed by sharp cubic edges and promotes asymmetric membrane stretching without full lysis. Together, these shape, magnetic, and surface effects explain why the C35 sample, despite its strong magnetic performance, induces erythrocyte deformation. Its high magnetic and magnetocrystalline anisotropies enhance hyperthermic heating but impair hemocompatibility, as repeated RBC deformation could compromise blood flow or accelerate splenic clearance [121-122]. Accordingly, although this sample exhibits the highest cellular uptake in SKOV3 ovarian cancer cells (for both 100 and 500 μg$mL^{-1}$) and reaches temperatures exceeding 40 °C in tumor-mimicking phantoms, its morphological impact on RBCs under physiologically relevant concentrations indicates it is unsuitable for IV administration. Instead, it is better suited for localized intratumoral injections, where direct blood contact is limited and precise thermal control is prioritized.

In contrast, the $Zn_{0.35}Mn_{0.25}Fe_{2.4}O_4$ nanoparticles, both cubic and spherical (C3 and S3) (**Fig. 6D&E; supplementary materials Fig. S5**), exhibited exceptional hemocompatibility and morphological stability even at the highest tested concentrations and time points. No RBC deformation or aggregation was observed after 1 h or 24 h, consistent with minimal erythrocyte membrane perturbation. This behavior is consistent with their smaller size, lower magnetic hardness, and stronger citrate-mediated electrostatic stabilization, which together would be expected to reduce interparticle association and limit disruptive contact with the erythrocyte membrane[123-124]. However, for $Co_{0.6}Fe_{2.4}O_4$ nanoparticles, both morphologies (C4 and S4) (**supplementary materials Fig. S3 & Fig. S4**) remained non-hemolytic (< 0.03%), but RBC morphology showed a clear shape-dependent response. C4 sample induced visible erythrocyte deformation at both 200 and 1000 μg $mL^{-1}$, with changes already evident after 1 h. The effect was dose-dependent, with noticeably milder deformation at 200 μg $mL^{-1}$ than at 1000 μg $mL^{-1}$. In contrast, spherical Co-ferrite particles (S4) preserved near-native biconcave morphology at 1 h at both concentrations but produced clear morphology disruption after 24 h at both 200 μg $L^{-1}$ and 1000 μg $mL^{-1}$, consistent with a delayed, exposure-time-dependent effect [125-126]. This divergence is consistent with Co-ferrite's higher magnetocrystalline anisotropy combined with faceted geometry in quasi-cubes, which enhances dipolar coupling and clustering as concentration increases and concentrates mechanical stress at particle edges during RBC contact. For the S4 sample, the delayed 24 h effect is consistent with cumulative lipid bilayer fatigue and progressive cytoskeletal remodeling arising from sustained particle-membrane contact over prolonged incubation [125, 127-130]. Thus, Co-ferrite exhibits acceptable hemolysis but a morphology liability that is strongest and earliest for quasi-cubic particles and increases with concentration.

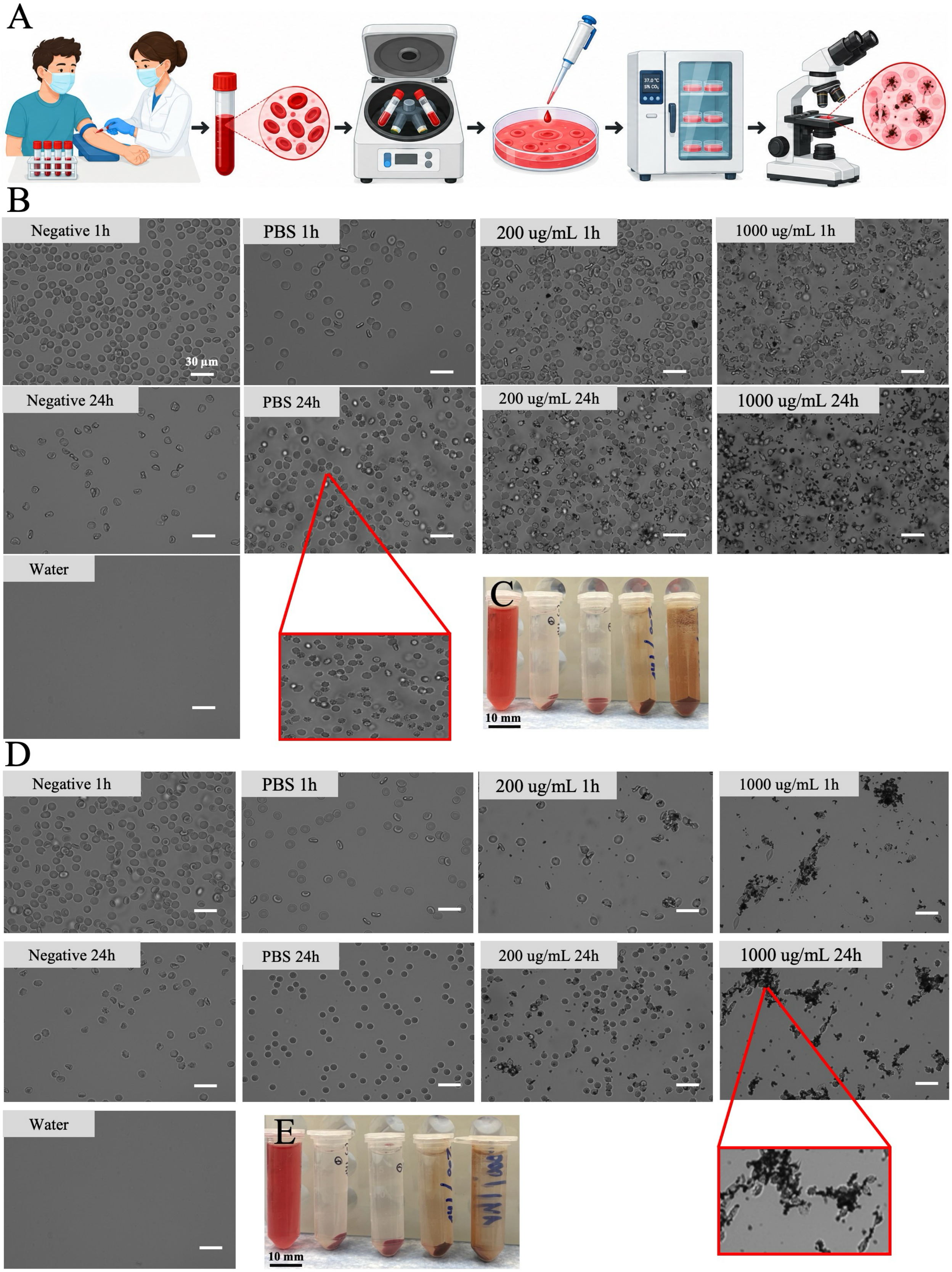

**Fig. 6.** Workflow of the hemolysis assay and qualitative morphological evaluation of red blood cells after nanoparticle exposure. (A) Schematic representation of the RBC incubation and imaging procedure. (B) and (D) are representative bright-field microscopy images of RBCs incubated with AS3, PBS, water, and C3 (B) and C35 (D) nanoparticle suspensions at 200 and 1000 µg/mL for 1 h and 24 h. Insets highlight representative burr cell morphology in PBS and cell aggregation at high nanoparticle concentration. (C) and (E) are representative

photographs of the corresponding suspensions after incubation. The observed morphological alterations were used as qualitative indicators of RBC membrane damage and blood compatibility. Schematic created using Canva.

Results on RBC counts showed no change in the size distribution of RBCs incubated with MNPs after either 1 h or 24 h, indicating that the samples did not alter the diameter of the cells, as shown in the representative plots in **Fig. 7A-D** and **supplementary materials Fig. S6(A-F)**. Furthermore, all sample cell counts (number of cells $mL^{-1}$) were within two standard deviations of their respective controls. The hemolytic activity of samples C35 and C3, assessed as the absorbance measured by U.V.-Vis spectroscopy at 1 hour and 24 hours in **Fig. 7E&F**, revealed that the supernatant from RBCs incubated with MNPs displayed higher absorbance values than those of the negative control (PBS) and the preservation solution (AS3); the same behavior holds for all other materials in the **supplementary materials Fig. S7(A-J) & Fig. S8(A-F)**. However, when the hemolysis percentage was calculated, lysis remained negligible for all samples. No coagulation, lysis, or cellular debris was observed. Overall, these results highlight that hemocompatibility is dictated not solely by shape but by the interplay of morphology, magnetic anisotropy, and surface chemistry. Large, high-moment C35 cubes induce RBC deformation even under non-hemolytic conditions, whereas small, magnetically softer C3 samples, despite their quasi-cubic geometry, remain fully hemocompatible. Therefore, while C35 nanoparticles achieve superior cellular uptake and hyperthermic heating, their RBC interactions render them unsuitable for IV use. Conversely, small C3 nanoparticles provide an ideal balance between magnetic performance and systemic safety, making them the most promising candidates for intravenous, MRI-integrated magnetic hyperthermia targeting ovarian tumors.

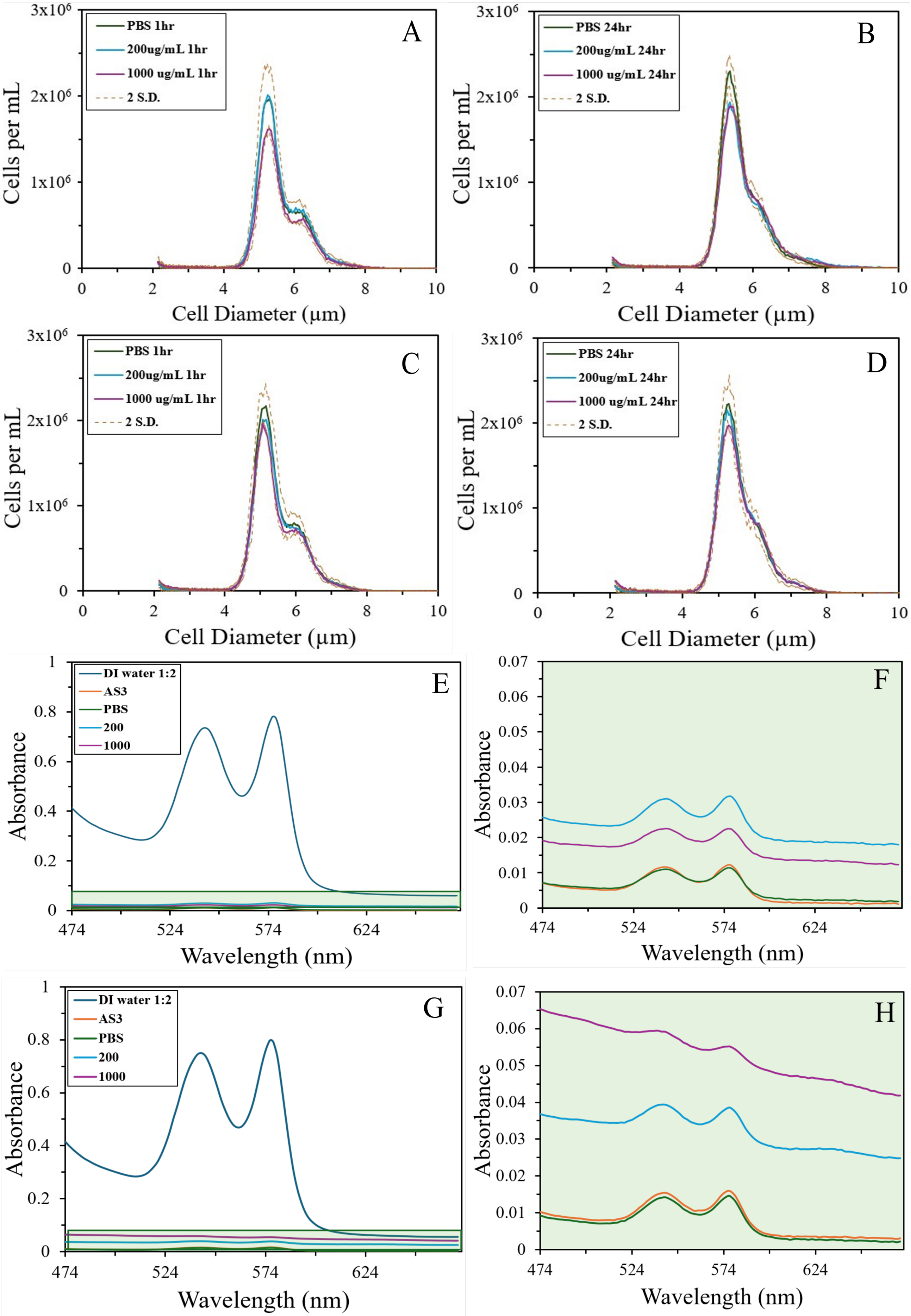
A
PBS 1hr
200ug/mL 1hr
1000 ug/mL 1hr
2 S.D.
Cells per mL
Cell Diameter (µm)
B
PBS 24hr
200ug/mL 24hr
1000 ug/mL 24hr
2 S.D.
C
D
E
DI water 1:2
AS3
PBS
200
1000
Absorbance
Wavelength (nm)
F
G
H

**Fig. 7.** Cell Counts of RBCs incubated with C35 (A, B) and C3 (C, D) MNPs evaluated after 1 h (A, C) and 24 h (B, D) at concentrations of 200 and 1000 μg $mL^{-1}$ in deionized water; dashed lines indicate the mean ± 2 S.D. (standard deviations). (E, G) U.V.-Vis absorption spectra for C35 (E) and C3 (F) after 1 h incubation time show minimal changes relative to the PBS control, while deionized water induces a strong hemoglobin release signal. (F, H) Magnified views of the hemoglobin absorption region for (E) and (G).

### 2.6. Effect of Magnetic Nanoparticles on Cell Viability in 3D Printed Perfusable Tumor Model

The 3D perfusable tumor model was used to evaluate the biocompatibility and cellular response of C35 and C3 nanoparticles under dynamic culture conditions that mimic key transport features of a vascularized tumor microenvironment. SKOV3 ovarian cancer cells were embedded in GelMA-gelatin (see the GelMA structure in **Fig. 7I**) tumor models containing a central perfusable microchannel, through which continuous medium flow (~16 μL $min^{-1}$) enabled convective delivery and supported nutrient and oxygen exchange throughout the construct (**Fig. 7A**). This perfusion channel acts as a vascular analog, enabling convective-diffusive transport and generating physiological shear stress comparable to peritoneal organs *in vivo* [131-134]. Such vascular-mimicking perfused models are used to capture key features of transport in tissues, including convective delivery in the channel and diffusion into the surrounding matrix, and therefore provide a practical platform for nanoparticle transport and nanoparticle-cell interaction studies under flow [135-138]. After 48 h of perfusion with nanoparticle suspensions (100, 300, and 500 μg $mL^{-1}$), Live/Dead imaging showed dose-dependent cytocompatibility differences between the two formulations. In the C35 constructs, the viability dropped from 87% (100 μg $mL^{-1}$) (**Fig. 8B**) to 76% (300 μg $mL^{-1}$) (**supplementary materials Fig. S10**) and 69% (500 μg $mL^{-1}$) (**Fig. 8C&D**). In contrast, the C3 constructs showed higher viability of 98% (**Fig. 8E**), 91% (**supplementary materials Fig. S10**), and 86% (**Fig. 8F&G**) across the same concentrations. The consistently lower viability observed for C35 at matched doses is plausibly consistent with size-shape-dependent particle-cell interactions under dynamic exposure, where larger, rigid, faceted particles can increase membrane contact and perturb cell attachment relative to smaller nanoparticles[139]. In addition, iron-oxide surfaces may contribute to low-level redox activity in biological media, potentially adding a sublethal oxidative-stress component during prolonged exposure [140-142]. Together, these effects provide a plausible explanation for the reduced viability observed for C35 cubes compared with the smaller C3 ferrite nanoparticles.

In contrast, the higher viability observed with the C3 sample under perfusion is consistent with its small size and more negative ζ-potential (-61.3 mV), which together indicate higher colloidal stability [143]. This increased stability likely reduced agglomeration and sedimentation, helping to maintain a more uniform delivered nanoparticle dose within the 3D model over prolonged exposure [144]. Finally, ferrite nano-bio interactions, including oxidative-stress responses, are known to depend on both surface chemistry and dopant composition. Thus, the C3 sample surface composition may reduce Fenton-like •OH generation relative to undoped C35, and the citrate-derived interface also provides a plausible basis for the favorable viability profile observed here [145]. Overall, these results show that both C35 and C3 nanoparticles are cytocompatible in a perfused 3D vascular-mimicking tumor construct over the tested conditions, while highlighting a reduction in viability for the larger C35. The perfused 3D model, therefore, provides a physiologically informed platform for screening nanoparticle cytocompatibility under dynamic exposure prior to *in vivo* studies.

### 2.7. MRI Contrast in 3D Tumor-Mimicking Phantoms

In 3D tumor-mimicking phantoms loaded with C3 and C35 samples, both formulations generated a negative MRI contrast consistent with the strong transverse relaxation and local field perturbations produced by ferrite/iron-oxide nanoparticles (**Fig. 8H**) [146-147]. Notably, the C35-loaded phantom exhibited a more pronounced and spatially extended signal hypointensity in comparison with C3, consistent with stronger susceptibility-induced T2 relaxation enhancement arising from steeper local field gradients [148]. This signal heterogeneity and distortion in the C35 sample are expected because magnetic inclusions with higher effective susceptibility and larger magnetic moment generate stronger local perturbations ($\Delta B_0$) and steeper microscopic field gradients, which accelerate intravoxel dephasing (T2 shortening) and broaden the spatial extent of signal loss[149]. Because C35 nanoparticles have larger magnetic core volumes, their increased magnetic moment per particle and their tendency to behave as a larger effective magnetic inclusion in a 3D construct can amplify local field inhomogeneity compared with the smaller C3 formulation [150-152]. In addition to contributing to signal heterogeneity, susceptibility can also cause spatially non-uniform signal distribution within the phantom. The local field gradients (ΔB0) generate intravoxel field variations that, when sufficiently steep, exceed the refocusing capacity of the 180° pulse, producing residual phase dispersion at voxel boundaries. This residual dephasing, combined with non-uniform particle distribution arising from C35's limited colloidal stability, produces spatially irregular T2 shortening, meaning signal loss is unevenly distributed across the phantom, and the nanoparticle distribution can appear heterogeneous and geometrically irregular [153-155]. Quantitative measurements in the tumor phantoms (**Fig. 8J**) showed that C3 produced a clearer, more concentration-dependent T2 signal attenuation. This is consistent with susceptibility-induced signal heterogeneity and spatially non-uniform T2 shortening in the C35 phantoms, which can produce irregular and unevenly distributed hypointense regions. Overall, the stronger field distortion observed for C35 is consistent with a susceptibility-dominated contrast regime in which particle magnetic moment and effective magnetic inclusion size (which is driven primarily by particle size and local dispersion state in the 3D matrix) govern the magnitude and spatial footprint of ΔB0, and hence the severity of signal heterogeneity and hypointense void extent relative to C3 [156-157]. Because pronounced signal heterogeneity and spatially irregular T2 shortening can make hypointense regions appear substantially larger than their true physical extent and can obscure true lesion or tumor boundaries, C35 formulations that generate such severe susceptibility artifacts are unlikely to be practical candidates for subsequent *in vivo* MRI-integrated MHT.

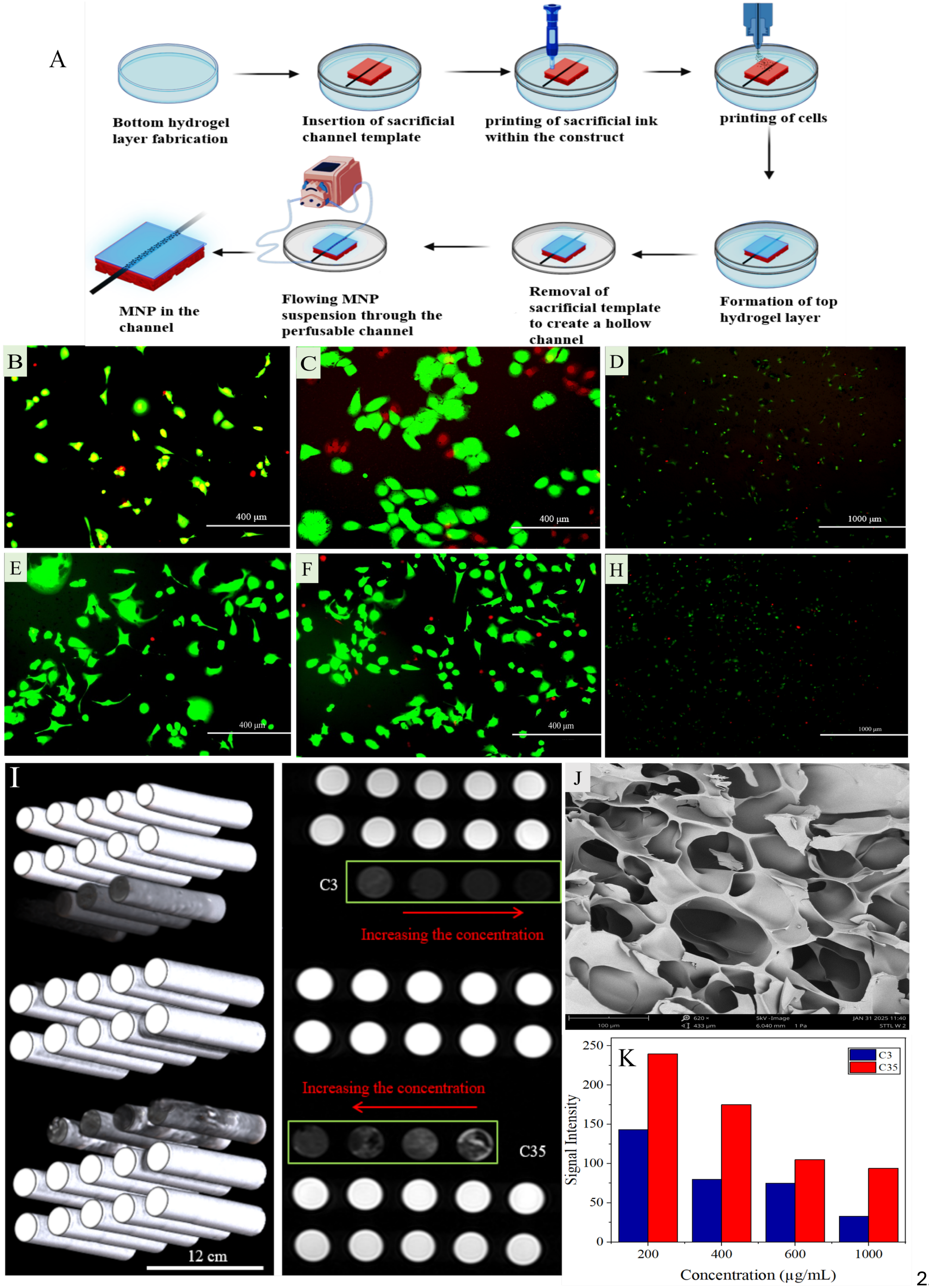
A
Bottom hydrogel layer fabrication
Insertion of sacrificial channel template
printing of sacrificial ink within the construct
printing of cells
Formation of top hydrogel layer
Removal of sacrificial template to create a hollow channel
Flowing MNP suspension through the perfusable channel
MNP in the channel
B
C
D
E
F
H
400 µm
1000 µm
I
12 cm
C3
Increasing the concentration
Increasing the concentration
C35
J
K
C3
C35
Signal Intensity
Concentration (µg/mL)
0
50
100
150
200
250
200
400
600
1000

**Fig. 8.** Perfused 3D vascular-mimicking tumor construct and MRI phantom evaluation of lead ferrite formulations. (A) Schematic illustration of the GelMA-gelatin tumor construct with a central perfusable microchannel and nanoparticle-containing medium flow. Representative Live/Dead images of SKOV3-laden constructs perfused with C35 at (B) 100 and (C) 500 μg $mL^{-1}$, respectively. Scale bar: 400 μm. (D) Higher-magnification Live/Dead image for C35 at 500 μg $mL^{-1}$. Scale bar: 1000 μm. Representative Live/Dead images of constructs perfused with C3 at (E) 100 and (F) 500 μg $mL^{-1}$, respectively. Scale bar: 400 μm. (G) Higher-magnification Live/Dead image for C3 at 500 μg $mL^{-1}$. Scale bar: 1000 μm. (H) T2-weighted MRI images of tumor-mimicking phantoms containing C3 and C35 at the 100, 200, 500, and 1000 μg $mL^{-1}$ concentrations. (I) SEM image of the GelMA microstructure. (J) T2 signal-intensity analysis for C3 and C35 phantoms versus concentration.

## 3. Conclusions

We developed a shape- and composition-controlled library of citrate-stabilized ferrite nanoparticles and systematically linked physicochemical properties to MRI contrast, magnetic hyperthermia performance, and biological safety in ovarian cancer-relevant models. Across clinically relevant AMF conditions, heating was strongly medium- and morphology-dependent: large $Fe_3O_4$ cubes (C35) delivered the highest thermal output but exhibited pronounced susceptibility-induced signal heterogeneity and spatially extended hypointensity in MRI and concentration-dependent RBC morphological deformation and reduced viability in perfused 3D cytocompatibility assays, whereas the smaller quasi-cubic Zn-Mn ferrite (C3) retained good heating efficiency under immobilized conditions while providing robust T2 contrast with minimal RBC morphology disruption and superior viability under perfusion. Live/Dead outcomes broadly tracked delivered thermal dose but were not strictly monotonic with bulk $\Delta T$, consistent with the role of local particle-cell association and microenvironmental constraints. Collectively, these results identify quasi-cubic Zn-Mn ferrite nanoparticles as the most promising balanced candidates for future evaluation toward systemically administered, MRI-integrated MHT, while positioning large $Fe_3O_4$ cubes as strong candidates for localized intratumoral hyperthermia where maximal heating is prioritized, and systemic exposure is limited.

## 4. Experimental Section

### 4.1. Materials

Ferric chloride hexahydrate ($FeCl_3 \cdot 6H_2O$, ≥98%), ferrous chloride tetrahydrate ($FeCl_2 \cdot 4H_2O$, ≥98%), zinc nitrate hexahydrate ($Zn(NO_3)_2 \cdot 6H_2O$, ≥98%), manganese nitrate tetrahydrate ($Mn(NO_3)_2 \cdot 4H_2O$, ≥98%), and cobalt nitrate hexahydrate ($Co(NO_3)_2 \cdot 6H_2O$, ≥98%) were purchased from Sigma–Aldrich (St. Louis, MO, USA). Sodium hydroxide (NaOH, pellets, ≥98%), ammonium hydroxide ($NH_4OH$), citric acid (≥99.5%), ethanol (200 proof), agar powder, sodium alginate, phosphate-buffered saline (PBS), and 0.25% trypsin solution were also obtained from Sigma–Aldrich (St. Louis, MO, USA) or Fisher Scientific (Waltham, MA, USA) as specified. All aqueous solutions were prepared using Deionized water (DI). Dulbecco's Modified Eagle Medium (DMEM; Sigma–Aldrich, St. Louis, MO, USA), fetal bovine serum (FBS; Corning, USA), and antibiotics (Corning, USA) were used for cell culture of SKOV3 ovarian cancer cells (ATCC, Manassas, VA, USA). Live/Dead Calcein-AM/EthD-1 viability kits were obtained from Biotium (Fremont, CA, USA). All chemicals were used without further purification.

### 4.2. Synthesis of Ferrite Nanoparticles

#### 4.2.1. Synthesis of Quasi-Cubic Ferrite Nanoparticles (Co-precipitation Method)

Small-sized quasi-cubic $Fe_3O_4$, $Zn_{0.3}Fe_{2.7}O_4$, $Zn_{0.35}Mn_{0.25}Fe_{2.4}O_4$, and $Co_{0.6}Fe_{2.4}O_4$ nanoparticles were synthesized via an alkaline co-precipitation route. In a typical procedure, $FeCl_3 \cdot 6H_2O$ and $FeCl_2 \cdot 4H_2O$ (molar ratio 1.9:1) were dissolved in 30 mL DI water under vigorous stirring at 35 °C. For doped ferrites, the corresponding metal nitrates ($Zn^{2+}$, $Mn^{2+}$, or $Co^{2+}$) were incorporated by substituting the appropriate fraction of $Fe^{2+}/Fe^{3+}$ precursors. A freshly prepared NaOH solution (2 M, 50 mL) was added dropwise to the reaction mixture until the pH reached approximately 11, resulting in the immediate formation of a black precipitate.

The suspension was kept at 35 °C for 1.5 h to complete nucleation and growth. After precipitation, an aqueous citric acid solution was introduced, and the mixture was stirred for an additional 30 min, allowing citrate ions to adsorb onto the nascent particle surfaces and impart electrostatic stabilization without fully passivating faceted growth. The nanoparticles were magnetically separated, washed five times with DI water and ethanol, and dried in a vacuum oven at 60 °C for 12 h.

**4.2.2. Synthesis of Spherical Ferrite Nanoparticles (Hydrothermal Method)**

Spherical $Fe_3O_4$, $Zn_{0.3}Fe_{2.7}O_4$, $Zn_{0.35}Mn_{0.25}Fe_{2.4}O_4$, and $Co_{0.6}Fe_{2.4}O_4$ nanoparticles were prepared using a hydrothermal method in which pH and temperature were tuned according to dopant chemistry. Metal precursors ($FeCl_3 \cdot 6H_2O$ and the corresponding nitrate salts) were dissolved in 25 mL DI water, followed by adjustment of the reaction pH to 9.5–10.5 using $NH_4OH$. The resulting suspension was briefly stirred at room temperature, after which the precipitate was collected by centrifugation and washed five times with deionized water to remove excess ammonium and free ions. The washed precipitate was then redispersed in deionized water, 6 mmol of citric acid was added, and the mixture was stirred for 30 min to allow citrate adsorption onto the particle surface, thereby providing electrostatic stabilization and improving colloidal dispersibility.

The solution was transferred to a Teflon-lined stainless-steel autoclave and heated at 180 °C-190 °C, depending on the composition. The autoclave was held at temperatures of 15 or 16 accordingly. After cooling to room temperature, the nanoparticles were magnetically collected, rinsed with DI water and acetone, and dried at 60 °C under vacuum.

**4.3. Characterization Techniques**

The structural, morphological, magnetic, and interfacial properties of the ferrite nanoparticles were comprehensively characterized using X-ray diffraction (XRD), electron microscopy, magnetometry, and colloidal analysis. After synthesis, all nanoparticle powders were dried in a vacuum oven for ≈ 6 h to obtain the final solid form. The crystal structure and phase purity were examined by powder XRD (Rigaku MiniFlex 6G diffractometer, Cu Kα, $\lambda = 1.5406$ Å), using a 1.25° slit and scanning over $2\theta = 20$-80° with a step size of 0.1°. Crystallite sizes were estimated from the full width at half-maximum (FWHM) of the main spinel reflections using the Scherrer equation, and lattice parameters were refined to probe subtle cation-distribution effects.

Nanoparticle morphology, shape uniformity, and core-size distributions were characterized by TEM (Hitachi 7650). For each sample, around 50 to 100 individual particles were manually segmented and measured using ImageJ software to obtain the mean core diameter and standard deviation, while high-magnification images were used to distinguish quasi-cubic from spherical morphologies. Colloidal behavior was assessed by dynamic light scattering (DLS, Microtrac Zetatrac, Model NPA152) in deionized water to determine hydrodynamic size and polydispersity. Zeta potential was measured using a Microtrac Stabino device at 25 °C, providing ζ-potential values that report on surface charge and electrostatic stabilization. Together, these measurements establish a direct link between crystal structure, size and shape, surface charge, and magnetic response in the engineered ferrite nanoparticle library.

Magnetic properties were measured using a Physical Property Measurement System (PPMS, Quantum Design) at room temperature under applied magnetic fields up to 15 kOe (150 mT). Magnetization curves were used to extract saturation magnetization (Ms), coercivity (Hc), and remanence (Mr). These values were correlated with particle size, morphology, and cation composition to interpret hyperthermia performance.

**4.3.1. Cell Culture and Cell Viability Analysis**

SKOV3 ovarian cell line (ATCC, Rockville, MD) was cultured according to the established protocol [25]. Cell viability following nanoparticle exposure was assessed using a fluorescence-based Live/Dead assay. SKOV3 ovarian cancer cells ($1 \times 10^5$ cell/mL) were seeded in standard 6-well plates and allowed to adhere overnight under standard culture conditions (37°C, 5 % $CO_2$). Nanoparticle suspensions were freshly prepared in complete culture medium, sterilized via 0.22 μm filtration, and added to the cells at final concentrations of 100 and 500 μg $mL^{-1}$. After 24 h incubation, the cells were rinsed twice with Dulbecco's Phosphate-Buffered Saline (DPBS) to remove unbound particles and stained with calcein-AM (2 μM) and ethidium homodimer-1 (EthD-1, 4 μM) for 30 min at room temperature in the dark. Fluorescence imaging was performed using an inverted fluorescence microscope (AMF4300, EVOS FL, USA). Live cells were identified by intracellular esterase-activated green fluorescence (calcein), while dead cells exhibited red nuclear staining (EthD-1).

**4.3.2. Cellular Uptake Analysis**

Cellular internalization of ferrite nanoparticles was assessed qualitatively using bright-field microscopy (AMF5000SV, EVOS FL, USA). SKOV3 ovarian cancer cells were cultured in 6-well plates at a density of $1 \times 10^5$ cells per well and incubated for 48 h to allow adhesion. Nanoparticle suspensions were prepared in FBS medium and added to each well at final concentrations of 100 and 500 μg $mL^{-1}$. For qualitative assessment, cells were fixed with 4% paraformaldehyde for 15 min.

**4.3.3. Hyperthermia Performance**

Magnetic hyperthermia experiments were performed using a MagneTherm system (nanoTherics Ltd.) equipped with water-cooled induction coils. For each measurement, the dried nanoparticle powder was redispersed in 1 mL of DI water or the desired biological medium and sonicated for 5 min in a bath sonicator to ensure homogeneous dispersion. The sample vial (1 mL) was then placed at the center of the coil, thermally insulated from the environment, and exposed to an alternating magnetic field with amplitude H = 30 mT and frequency f = 101 kHz. The real-time temperature of the suspension was monitored using an optical fiber thermometer inserted into the center of the sample, and temperature-time curves were recorded under continuous AMF exposure.

Hyperthermia tests were carried out in several media (total volume 1 mL), including FBS-supplemented SKOV3 cell culture medium, and tumor-mimicking agar constructs containing SKOV3 cells. Agar phantoms were prepared by dissolving 2 wt% agar in DI water, heating to 90-95 °C with magnetic stirring until optically clear, then cooling to ~50 °C. The warm agar solution was then combined with SKOV3 cell-incubated MNP suspensions to reach the final volume, gently mixed to avoid bubble formation, transferred into hyperthermia vials, and allowed gel at 4 °C for 30 min, thereby forming SKOV3 tumor-mimicking agar phantoms for hyperthermia measurements.

The SAR was determined from the initial slope of the temperature rise using the standard initial-slope method:

$$\mathrm{SAR} = \frac{m s_s\, C_{\mathrm{susp}}}{m_{\mathrm{MNP}}} \left(\frac{dT}{dt}\right)_{t\to 0},$$

where $C_{\mathrm{susp}}$ is the effective specific heat capacity of the suspension (J $g^{-1}$ $K^{-1}$), $m_{\mathrm{MNP}}$ is the mass of nanoparticles (g) in the 1 mL sample, and $(\mathrm{d}T/\mathrm{d}t)_{t\to 0}$ is the initial linear slope of the temperature-time curve (K $s^{-1}$), typically evaluated over the first 15-60 s. For convenience, the heating efficiency can also be expressed using the ILP, which normalizes SAR by the applied field frequency and amplitude:

$$\mathrm{ILP} = \frac{\mathrm{SAR}}{f.H^2} \left(\frac{\Delta T}{\Delta t}\right),$$

where f is the frequency of the alternative magnetic field, and H is the amplitude of the magnetic field.

**4.3.4. Hemolysis Performance**

Nanoparticle stock solutions were prepared by dispersing 16.0 mg of dried nanoparticles (C35, C3, C4, S3, and S4) in 8 mL of phosphate-buffered saline or PBS (Thermo Fisher Scientific, USA, Waltham, MA, USA), resulting in a concentration of 2.0 mg MNPs $mL^{-1}$. The suspensions were sonicated to achieve uniform dispersion and minimize agglomeration. Working solutions at 200 μg $mL^{-1}$ and 1000 μg $mL^{-1}$ were freshly diluted from the stock in PBS immediately before incubation with red blood cells. Human whole blood was collected from volunteers into EDTA tubes, with informed consent and approval from the Texas Tech University Health Science Center IRB (Protocol L22–L274). RBCs were isolated by repeated centrifugation and washing, then incubated with nanoparticle suspensions at 37 °C for either 1 or 24 hours under static conditions. PBS-treated RBCs served as negative controls, and for further comparison, RBCs were incubated in the additive solution AS-3, which maintains cell integrity during storage. After incubation, cell counts, hemolysis percentages, and microscopy images were obtained to assess nanoparticle biocompatibility. Hemolysis was quantified using an Agilent Cary 60 UV-Vis spectrophotometer (Agilent Technologies, Santa Clara, CA, USA) following the Winterbourn method for hemoglobin determination[158]. Hemolysis (%) was calculated as ([Free Hbsample] - [Free Hb(neg. cnt.)]) / ([Total Hb(pos. cnt.)] – [Free Hb(neg. cnt.)]) × 100%, where PBS and deionized water-treated RBCs served as negative and positive controls, respectively. RBC counts and volume distributions were measured with Multisizer 4e Coulter Counter (Beckman Coulter, CA, USA). Cell morphology was evaluated by optical microscopy using an EOVS M5000 imaging system (Thermo Fisher Scientific, USA, Waltham, MA, USA). Three biological replicates and three technical replicates were done for each test.

#### 4.3.5. T2-Weighted MRI of Tumor-Mimicking Agar Phantoms

The MRI contrast performance of the ferrite nanoparticles was evaluated in ovarian tumor-mimicking agar phantoms using a clinical 3.0 T MRI scanner with a head coil for signal reception. T2-weighted images were acquired using a turbo spin-echo sequence with the following parameters: TR = 3200 ms, TE = 409 ms, field of view (FOV) = 218 mm, slice thickness = 5.0 mm, and voxel size = 0.4 × 0.4 × 5.0 mm

#### 4.3.6. Cell Cytocompatibility Analysis in Perfusable 3D Tumor Model

Gelatin methacrylate (GelMA) and gelatin were employed as extracellular matrix (ECM) analogs for the fabrication of the three-dimensional (3D) tumor constructs, owing to their intrinsic cell-adhesive motifs that promote cellular attachment, spreading, and proliferation. GelMA was synthesized from gelatin following a previously established protocol. Ink preparation and fabrication of the perfusable, sandwich-structured 3D ovarian tumor model was carried out in accordance with reported methodologies. Briefly, a 6% (w/v) GelMA-gelatin hydrogel mixture was manually cast into a Polydimethylsiloxane (PDMS) mold to form a bottom layer of approximately 1 mm in thickness. After inducing rapid thermal gelation at 4 °C, a 500 μm diameter straight sacrificial filament of Pluronic F-127 was deposited using an extrusion-based 3D bioprinter (TissueStart, Tissuelabs, Switzerland) to define the future perfusion channel. Cell-laden bioink droplets ($1 \times 10^6$ cells/mL) were subsequently inkjet-printed adjacent to the Pluronic filament to spatially localize the tumor cells near the channel interface. Following cell deposition, an additional 6% GelMA-gelatin layer was applied to encapsulate the structure, creating a sandwich-like 3D tumor construct. The Pluronic F-127 was then liquefied by cooling to 4 °C and carefully evacuated using a syringe, resulting in the formation of a hollow, perfusable microchannel within the engineered tumor model. The finalized tumor construct was achieved through ultraviolet (UV)-mediated photocrosslinking of the hydrogel network. A peristaltic pump was connected to the tumor contract for perfusion of the cell culture medium containing nanoparticles. For MNPs cytocompatibility assays, the channel was perfused with SKOV3 culture medium containing C35 and C3 nanoparticles at defined concentrations (100, 300 and 500 μg $mL^{-1}$). The final tumor constructs were maintained at 37 °C with continuous medium flow (16 μL $min^{-1}$). After the desired perfusion period (24-48 h), cell viability within the 3D tumor region was assessed using a Live/Dead fluorescence assay, allowing direct comparison of SKOV3 viability in perfused conditions between $Fe_3O_4$ and Zn-Mn ferrite nanoparticle exposure.

#### 4.3.7. Scanning Electron Microscopy Imaging of The Hydrogel

To assess the hydrogel microstructure, constructs were first perfused with DPBS for 3 h to flush out residual culture medium and cellular debris. The samples were then frozen at -80 °C overnight and subsequently lyophilized for 3 days using a freeze dryer (Labconco, USA). Before scanning electron microscopy (SEM), the dried hydrogels were sputter-coated with gold (Ted Pella Inc., USA; 0.1 mA/mbar, 30 s, room temperature). Microstructural features were imaged on a Phenom Pharos SEM (Thermo Fisher Scientific, USA), and pore architecture and morphology were quantitatively analyzed using ImageJ software.

**Supplementary Materials**

Supplementary materials include:

S1. EDX spectra of quasi-cubic ferrite nanoparticles confirming elemental composition and citrate surface coating

S2. EDX spectra of spherical ferrite nanoparticles confirming elemental composition and citrate surface coating

S3. Optical microscopy images of RBCs incubated with S4 at 200 and 1000 µg $mL^{-1}$ for 1 h and 24 h, alongside negative (AS3) and PBS controls

S4. Optical microscopy images of RBCs incubated with C4 quasi-cubic Co-ferrite nanoparticles at 200 and 1000 µg $mL^{-1}$ for 1 h and 24 h

S5. Optical microscopy images of RBCs incubated with S3 spherical Zn-Mn ferrite nanoparticles at 200 and 1000 µg mL-1 for 1 h and 24 h

S6. RBC size distribution profiles of cells incubated with S4, C4, and S3 nanoparticles at 200 and 1000 µg $mL^{-1}$ for 1 h and 24 h relative to PBS control and ±2 SD (standard deviations) boundaries

S7. UV-Vis absorption spectra of supernatants collected after 24 h incubation of RBCs with C35, C3, C4, S4, and S3 nanoparticles at 200 and 1000 µg $mL^{-1}$

S8. UV-Vis absorption spectra of supernatants collected after 24 h incubation of RBCs with C35, C3, C4, S4, and S3 nanoparticles at 200 and 1000 µg $mL^{-1}$, UV-Vis absorption spectra of supernatants collected after 1 h incubation of RBCs with C4, S4, and S3 nanoparticles at 200 and 1000 µg $mL^{-1}$

S9. Bright-field microscopy images of SKOV3 cells after 48 h incubation with C35 and C3 nanoparticles at 500 µg $mL^{-1}$

S10. Live/Dead fluorescence images of SKOV3 cells in 3D perfusable GelMA tumor constructs after 48 h perfusion with C35 (A, C) and C3 (B, D) nanoparticles at 300 µg $mL^{-1}$

**Funding**

Research reported in this publication was supported by the National Institute Of Biomedical Imaging And Bioengineering of the National Institutes of Health under Award Number R03EB036435, the National Institute Of General Medical Sciences of the National Institutes of Health under Award Number R16GM158539, the National Institute Of Allergy And Infectious Diseases of the National Institutes of Health under Award Number R03AI188351, and the National Heart, Lung, And Blood Institute of the National Institutes of Health under Award Number R15HL181720. The content is solely the responsibility of the authors and does not necessarily represent the official views of the National Institutes of Health. J.G.-P. acknowledges support from The Welch Foundation under Grant Number D-2236-20250403 and the Cancer Prevention & Research Institute of Texas under Grant Number RP250634. K.W. acknowledges support from the U.S. National Science Foundation under award No. 2630047. Any opinions, findings and conclusions or recommendations expressed in this material are those of the author(s) and do not necessarily reflect the views of the U.S. National Science Foundation. B.R. and K.M.P.G. acknowledge the Distinguished Graduate Student Assistantships (DGSA) supported by Texas Tech University.

**Author Contributions**

B.R. and K.W. conceived the idea and designed the study. B.R. synthesized and characterized all ferrite nanoparticle formulations, performed all magnetic hyperthermia and MRI phantom experiments, analyzed and interpreted all experimental data, and wrote the manuscript. Md.S. and C.X. carried out the cellular uptake imaging and cell viability assays before and after AMF treatment and wrote the related part of the manuscript. S.M. contributed to nanoparticle physicochemical and magnetic characterization and prepared the schematic illustrations. K.M.P.G., N.T.L.T., B.R., and J.G.-P. performed hemocompatibility experiments, including hemolysis, RBC morphology, and blood cell count measurements. K.M.P.G. and J.G.-P. performed Coulter counter analysis and UV-Vis spectrophotometry data acquisition and plotted the corresponding datasets. B.R. and K.W. led the writing of the manuscript. Md.S., K.M.P.G., and J.G.-P. supported the writing of the manuscript. K.W. reviewed and proofread the manuscript.

**Conflicts of Interest**

The authors declare no conflicts of interest.

**Data Availability Statement**

Data supporting this study are openly available in Zenodo at: https://doi.org/10.5281/zenodo.22716018

**Author ORCID iDs:**

Bahareh Rezaei: 0000-0003-3446-3559
Md Shahriar: 0000-0003-4895-2601
Shahriar Mostufa: 0000-0002-3326-4817
Karla Mercedes Paz González: 0009-0006-6974-0594
Nguyen Thuy Linh Tran: 0009-0002-0130-4826
Changxue Xu: 0000-0002-5323-595X
Jenifer Gómez-Pastora: 0000-0002-5157-4130
Kai Wu: 0000-0002-9444-6112

**TOC Graphic:**

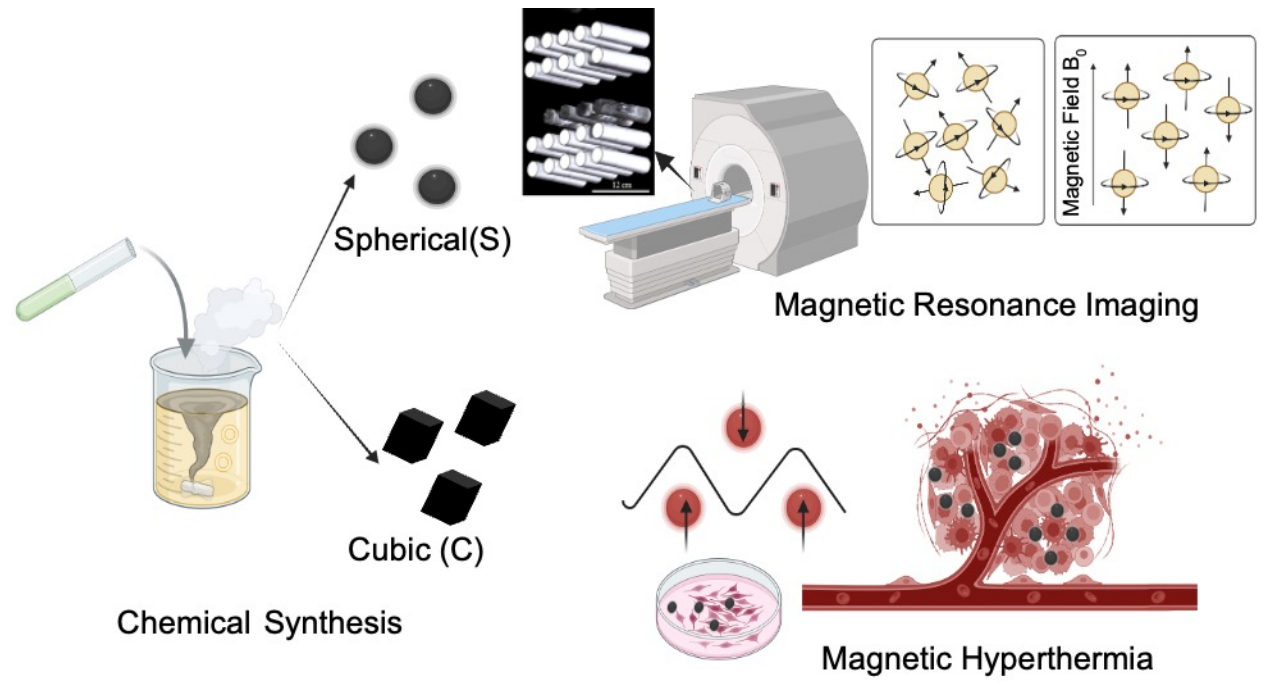

# Supplementary Materials

## Shape- and Cation-Engineered Ferrite Nanoparticles for Enhanced Theranostic Performance in Ovarian Cancer Tumor-Mimicking Phantom

Bahareh Rezaei[1], Md Shahriar[2], Shahriar Mostufa[1], Karla Mercedes Paz González[3], Nguyen Thuy Linh Tran[3], Changxue Xu[2], Jenifer Gómez-Pastora[3], Kai Wu[1,4,*]

[1]Department of Electrical and Computer Engineering, Texas Tech University, Lubbock, TX 79409, USA

[2]Department of Industrial, Manufacturing, and Systems Engineering, Texas Tech University, Lubbock, TX 79409, USA

[3]Department of Chemical Engineering, Texas Tech University, Lubbock, TX 79409, USA

[4]Department of Physics, University of South Florida, Tampa, FL 33620, USA

*Corresponding Author: kaiwu@usf.edu (K.W.)

**S1. EDX spectra of quasi-cubic ferrite nanoparticles confirming elemental composition and citrate surface coating**

**Figure S1** presents the EDX spectra of selected quasi-cubic ferrite nanoparticles (C1, C2, and C4). All spectra show characteristic peaks for O and Fe, consistent with the spinel ferrite core, alongside a detectable C signal confirming the presence of an organic surface layer attributable to citrate coating. The Zn signal detected in C2 and the Co signal in C4 confirm successful incorporation of the respective dopant cations into the ferrite structure. The Cu signal present in all spectra originates from the TEM copper support grid and is not attributed to the nanoparticle composition.

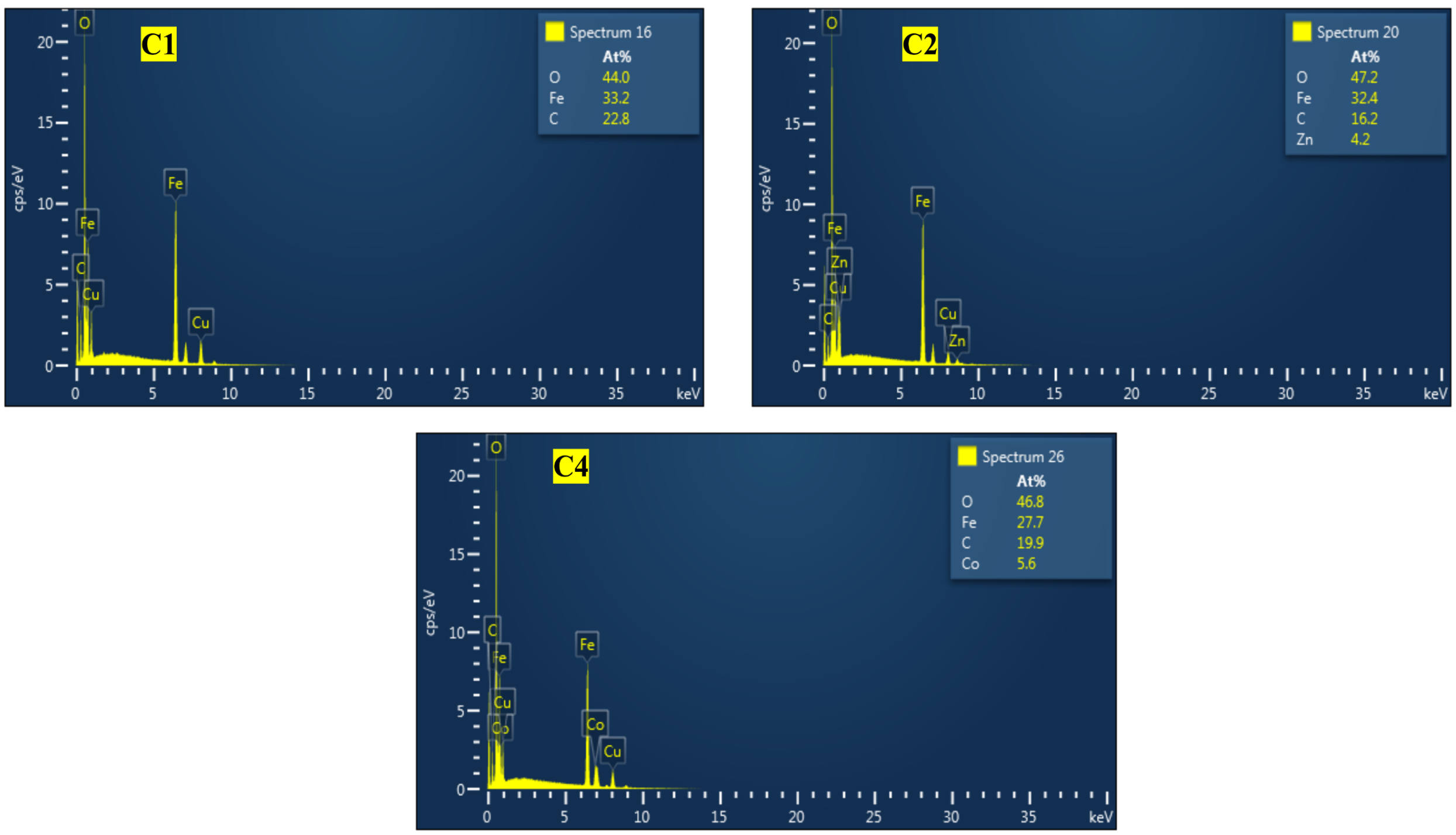

**Figure S1.** EDX spectra of quasi-cubic ferrite nanoparticles C1, C2, and C4, confirming elemental composition and presence of carbon signal consistent with citrate surface coating.

**S2. EDX spectra of spherical ferrite nanoparticles confirming elemental composition and citrate surface coating**

**Figure S2** presents the EDX spectra of the spherical ferrite nanoparticles (S1-S4). All spectra show characteristic O and Fe peaks consistent with the spinel ferrite core, alongside a detectable C signal confirming citrate surface coating. The Zn signal in S2, the Zn and Mn signals in S3, and the Co signal in S4 confirm successful incorporation of the respective dopant cations into each ferrite composition. The Cu signal originates from the TEM copper support grid and is not attributed to the nanoparticle composition.

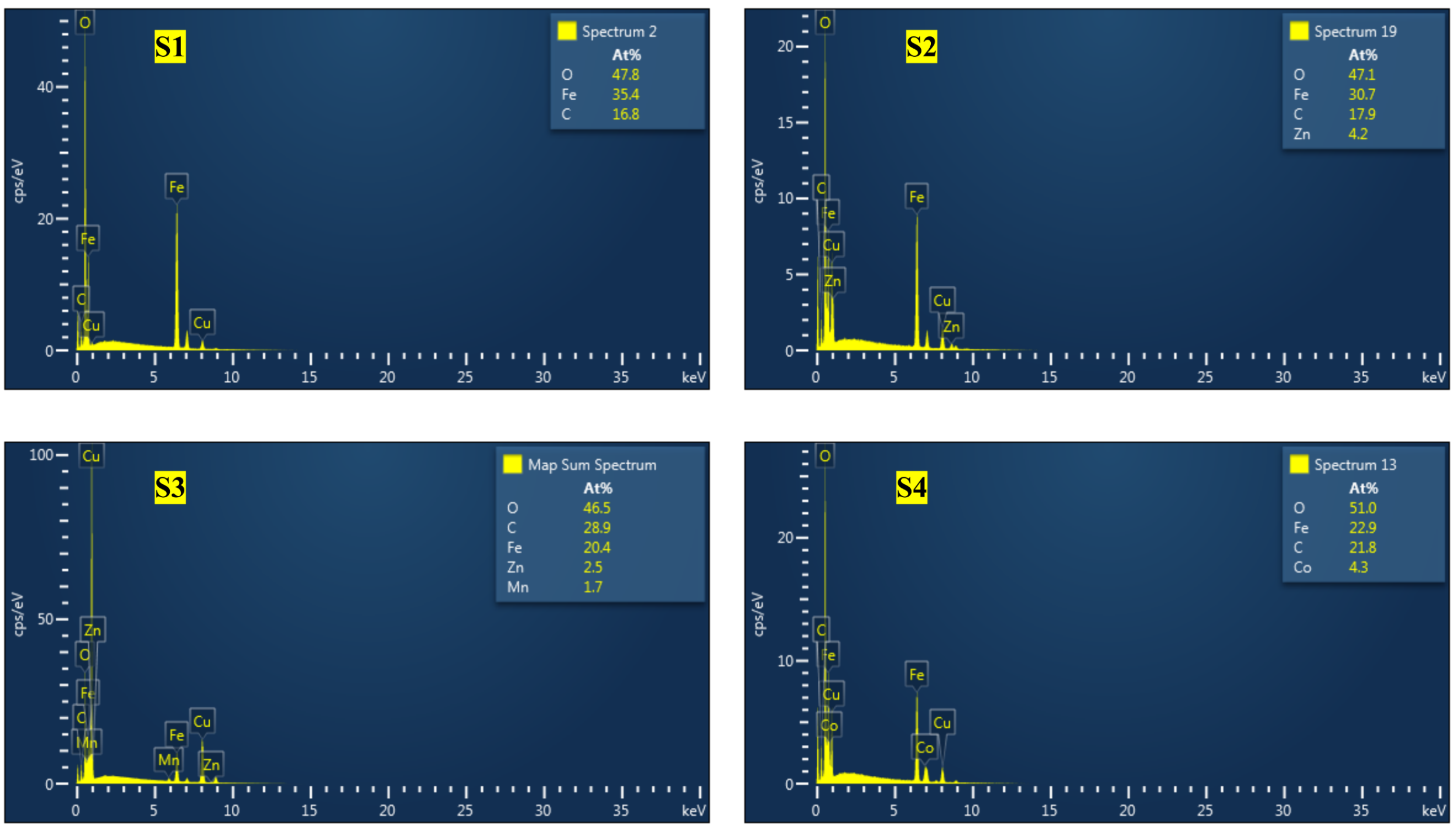

**Figure S2.** EDX spectra of spherical ferrite nanoparticles confirming elemental composition and citrate surface coating.

**S3. Optical microscopy images of RBCs incubated with S4 at 200 and 1000 µg mL$^{-1}$ for 1 h and 24 h, alongside negative (AS3) and PBS controls**

**Figure S3** shows optical microscopy images of RBCs incubated with S4 nanoparticles at 200 and 1000 µg mL$^{-1}$ for 1 h and 24 h. AS3 and PBS controls display normal biconcave disc morphology at both time points. At 1 h, RBCs incubated with S4 at both concentrations preserved normal morphology, consistent with minimal acute membrane perturbation. However, at 24 h, clear morphological disruption was evident at both concentrations, consistent with the delayed, exposure-time-dependent effect attributed to cumulative lipid bilayer fatigue arising from sustained particle-membrane contact over prolonged incubation in protein-free PBS. The dark aggregates visible at 1000 µg mL$^{-1}$ are consistent with nanoparticle clustering at high concentration.

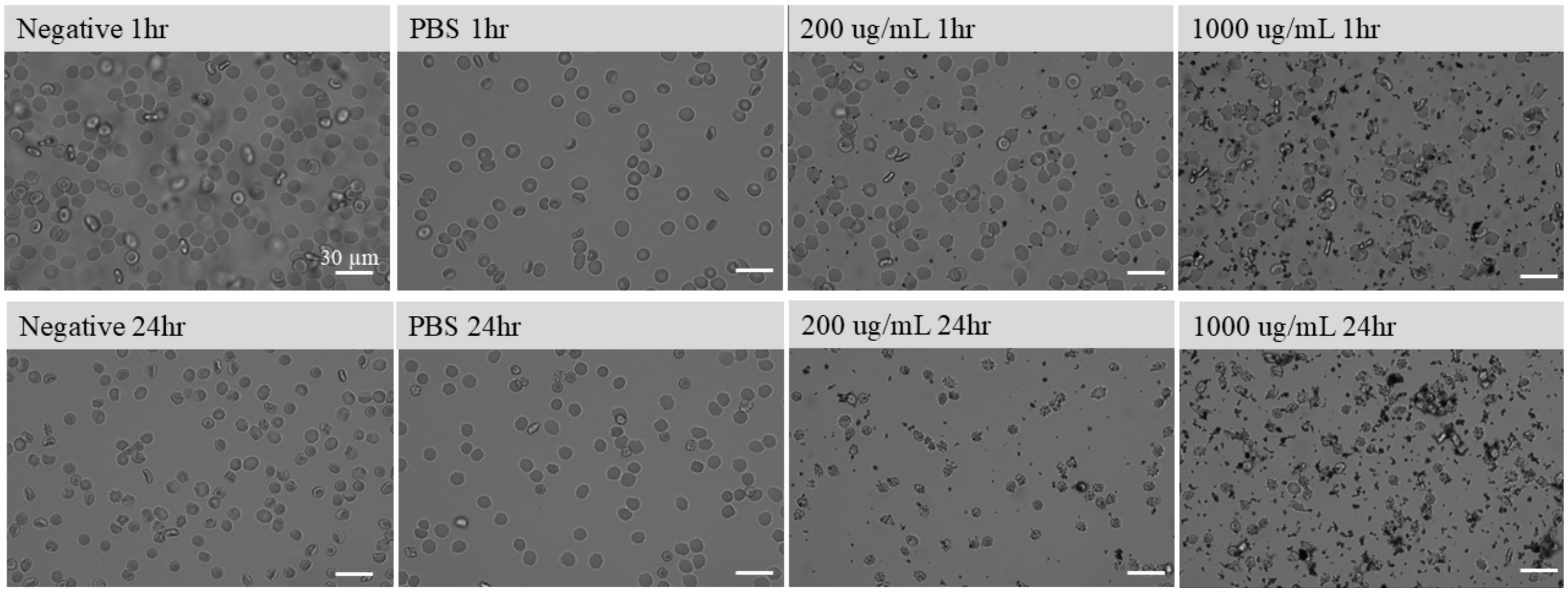

**Figure S3.** Bright-field microscopy images of RBCs incubated with Negative (AS3), PBS, water, and S4 sample suspensions at 200 and 1000 µg mL$^{-1}$ for 1 h and 24 h.

## S4. Optical microscopy images of RBCs incubated with C4 quasi-cubic Co-ferrite nanoparticles at 200 and 1000 µg mL$^{-1}$ for 1 h and 24 h

**Figure S4** shows optical microscopy images of RBCs incubated with C4 nanoparticles at 200 and 1000 µg mL$^{-1}$ for 1 h and 24 h. Negative (AS3) and PBS controls display normal biconcave morphology at both time points. In contrast to S4, C4 induced visible erythrocyte deformation at 1 h at both concentrations, with the effect being dose-dependent; morphological changes were noticeably more pronounced at 1000 µg mL$^{-1}$ than at 200 µg mL$^{-1}$. At 24 h, deformation was more extensive across both concentrations, consistent with the combined effects of Co-ferrite's high magnetocrystalline anisotropy promoting dipolar clustering and the concentration of mechanical stress at quasi-cubic particle edges during RBC contact. The dark aggregates visible, particularly at 1000 µg mL$^{-1}$, are consistent with concentration-dependent nanoparticle clustering

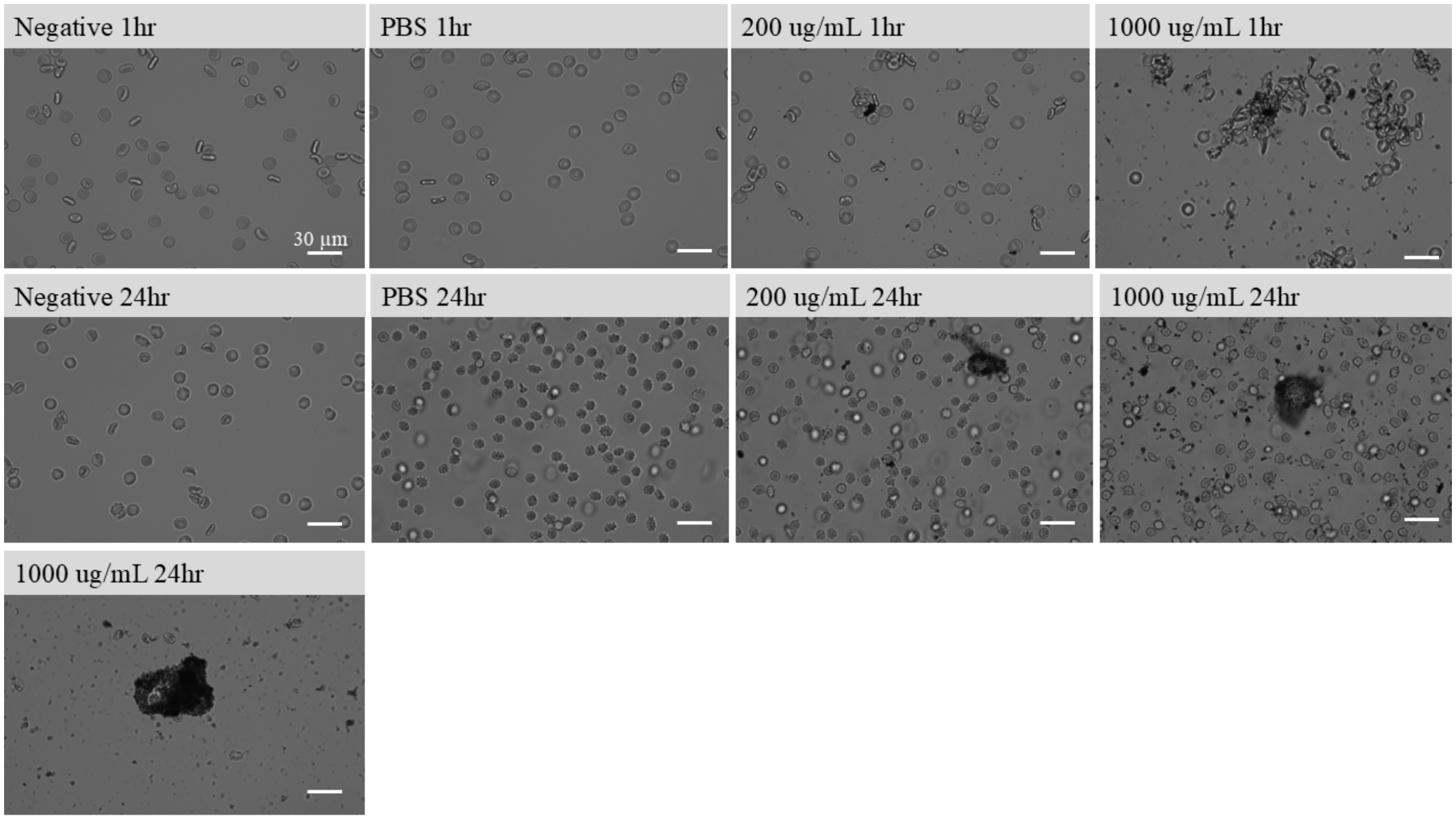

**Figure S4.** Bright-field microscopy images of RBCs incubated with negative (AS3), PBS, and C4 sample suspensions at 200 and 1000 µg mL$^{-1}$ for 1 h and 24 h.

**S5. Optical microscopy images of RBCs incubated with S3 spherical Zn-Mn ferrite nanoparticles at 200 and 1000 μg mL$^{-1}$ for 1 h and 24 h**

**Figure S5** shows optical microscopy images of RBCs incubated with nanoparticles at 200 and 1000 μg mL$^{-1}$ for 1 h and 24 h. AS3 and PBS controls display normal biconcave disc morphology at both time points. RBCs incubated with S3 retained near-native morphology across all tested concentrations and time points, with no visible deformation or aggregation observed at either 1 h or 24 h. The dark aggregates visible at 1000 μg mL$^{-1}$ are consistent with nanoparticle clustering at high concentration rather than RBC morphological changes, and no membrane disruption or lysis was detected.

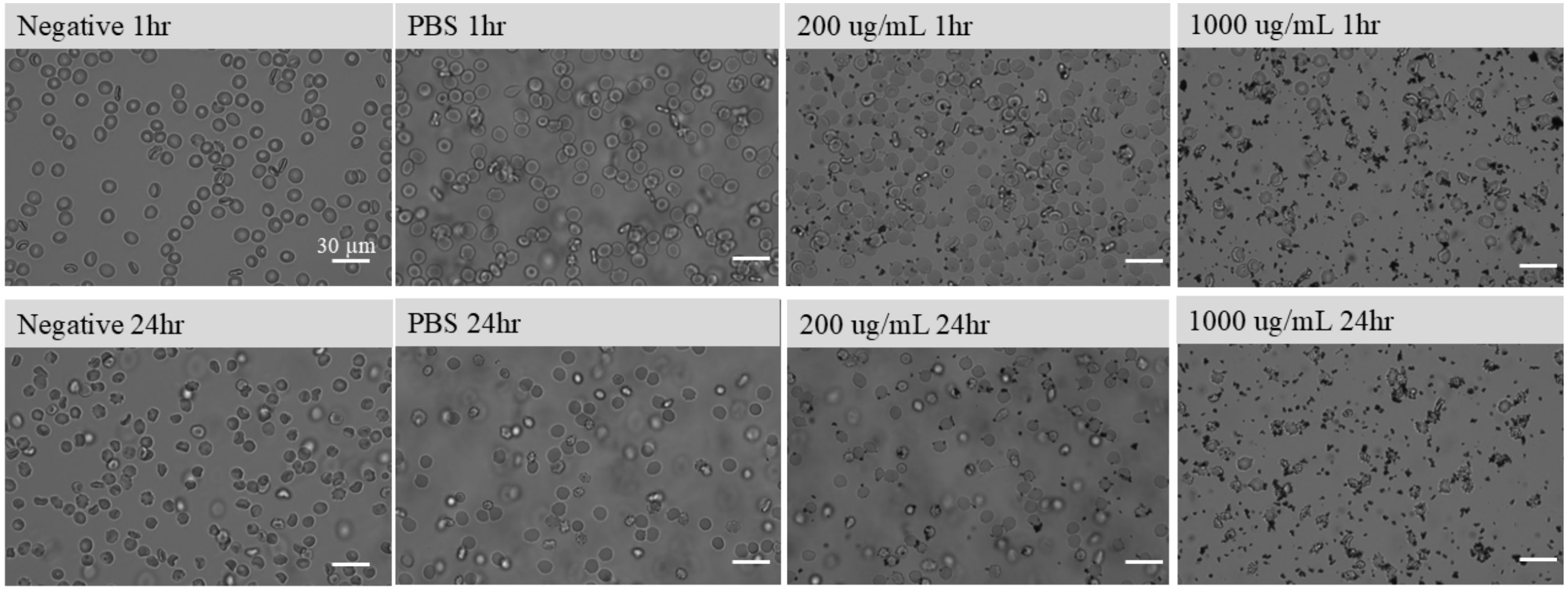

**Figure S5.** Bright-field microscopy images of RBCs incubated with negative (AS3), PBS, and S3 sample suspensions at 200 and 1000 μg mL$^{-1}$ for 1 h and 24 h.

**S6. RBC size distribution profiles of cells incubated with S4, C4, and S3 nanoparticles at 200 and 1000 µg mL$^{-1}$ for 1 h and 24 h relative to PBS control and ±2 SD (standard deviation) boundaries**

**Figure S6** shows the RBC size distribution profiles for cells incubated with S4, C4, and S3 nanoparticles at 200 and 1000 µg mL$^{-1}$ for 1 h (A, C, E) and 24 h (B, D, F). At both time points and concentrations, the size distribution curves overlap closely with the PBS control and remain within the ±2 SD boundary, indicating that these nanoparticles did not alter RBC diameter or volume distribution. The peak cell diameter remains consistently at approximately 5 µm across all conditions, confirming the absence of MNP-induced cell swelling, shrinkage, or fragmentation despite the morphological deformation observed by optical microscopy at 24 h.

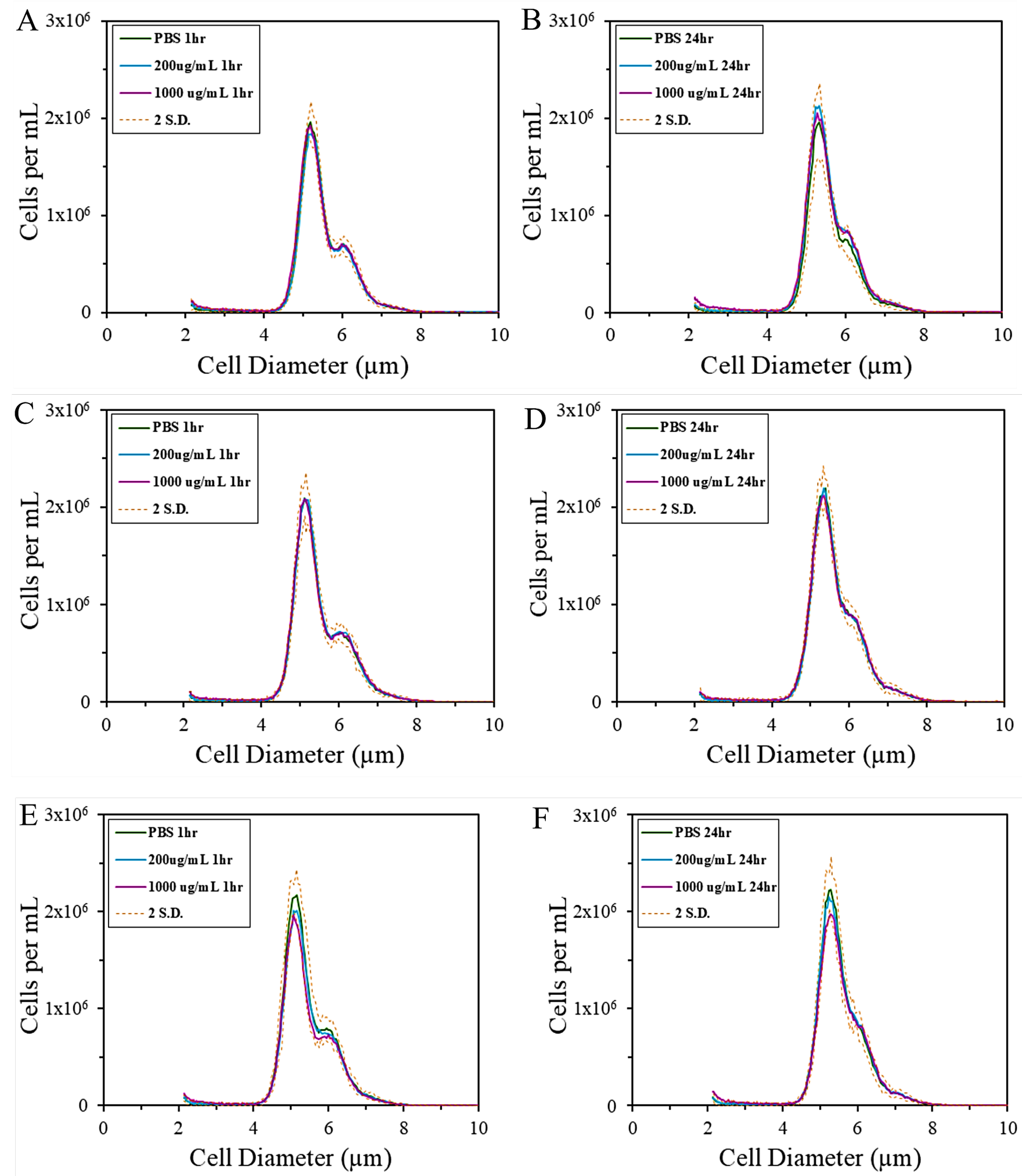

**Figure S6.** RBC size distribution profiles of cells incubated with S4 (A, B), C4 (C, D), And S3 (E, F) nanoparticles at 200 and 1000 µg mL$^{-1}$ for 1 h (A, C, E) and 24 h (B, D, F).

**S7. UV-Vis absorption spectra of supernatants collected after 24 h incubation of RBCs with C35, C3, C4, S4, and S3 nanoparticles at 200 and 1000 µg mL$^{-1}$**

**Figure S7** presents the UV-Vis absorption spectra of supernatants collected after 24 h incubation of RBCs with C35, C3, C4, S4, and S3 nanoparticles at 200 and 1000 µg mL$^{-1}$, alongside DI water, negative (AS3), and PBS controls. The positive control displays the characteristic oxyhemoglobin double-peak signature at approximately 541 and 577 nm with high absorbance, confirming complete hemolysis as a reference. However, Calculated hemolysis percentages remained negligible for both concentrations, confirming non-hemolytic behavior.

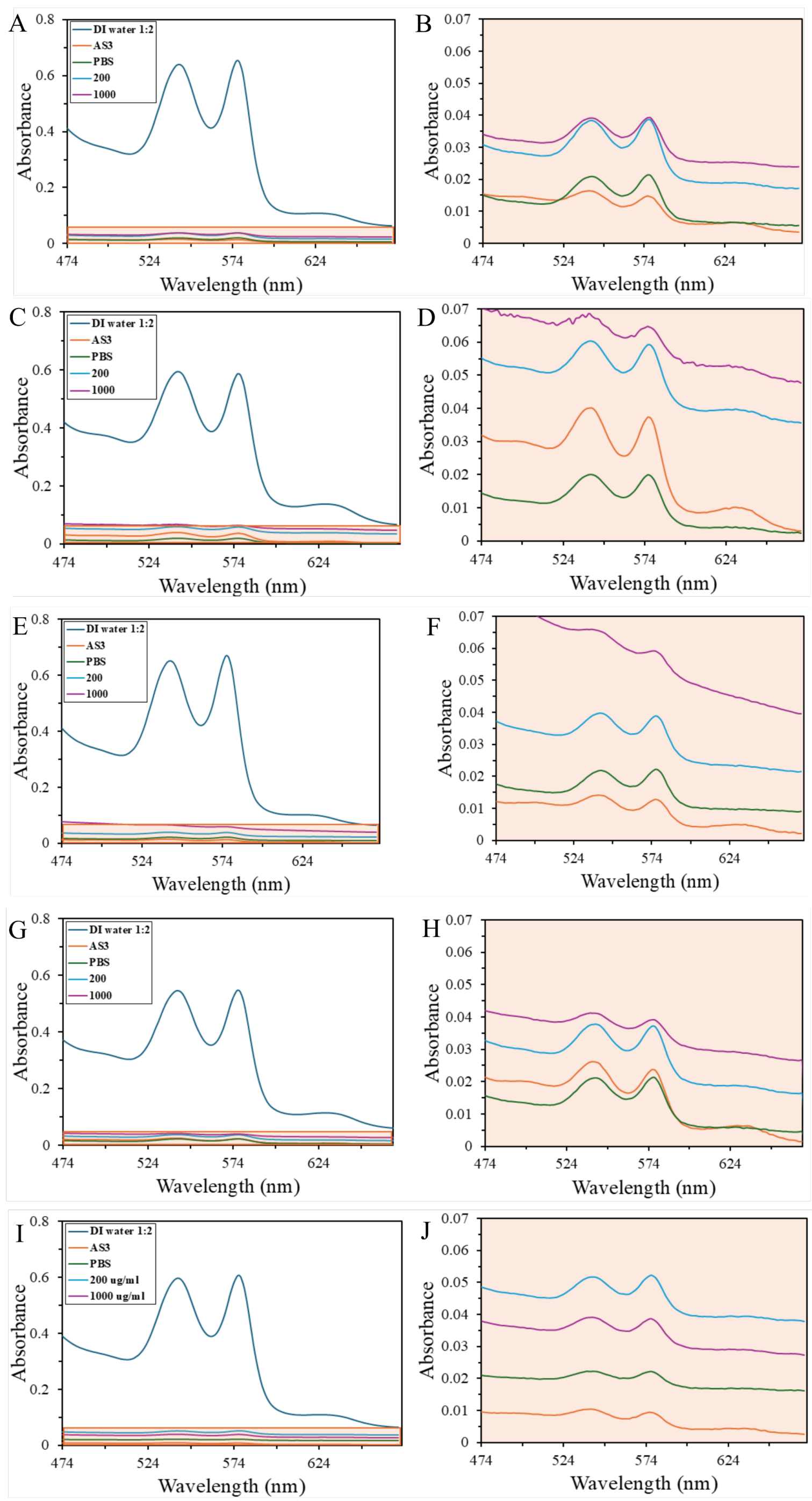

**Figure S7.** UV-Vis absorption spectra for C35 (A), C3 (C), C4 (E), S4 (G), and S3(I) after 24 h incubation. UV-Vis absorption spectra (A, C, E, G, and I)) and magnified hemoglobin region (B, D, F, H, and J) of supernatants collected after 24 h incubation of RBCs with different samples at 200 and 1000 µg mL$^{-1}$.

**S8. UV-Vis absorption spectra of supernatants collected after 1 h incubation of RBCs with C4, S4, and S3 nanoparticles at 200 and 1000 µg mL$^{-1}$**

**Figure S8** presents the UV-Vis absorption spectra of supernatants from RBCs incubated with C4, S4, and S3 nanoparticles at 200 and 1000 µg mL$^{-1}$ for 1 h. Despite visible erythrocyte deformation observed at 1 h by optical microscopy, calculated hemolysis percentages remained negligible at both concentrations, confirming that the morphological changes reflect mechanical membrane stress rather than lysis.

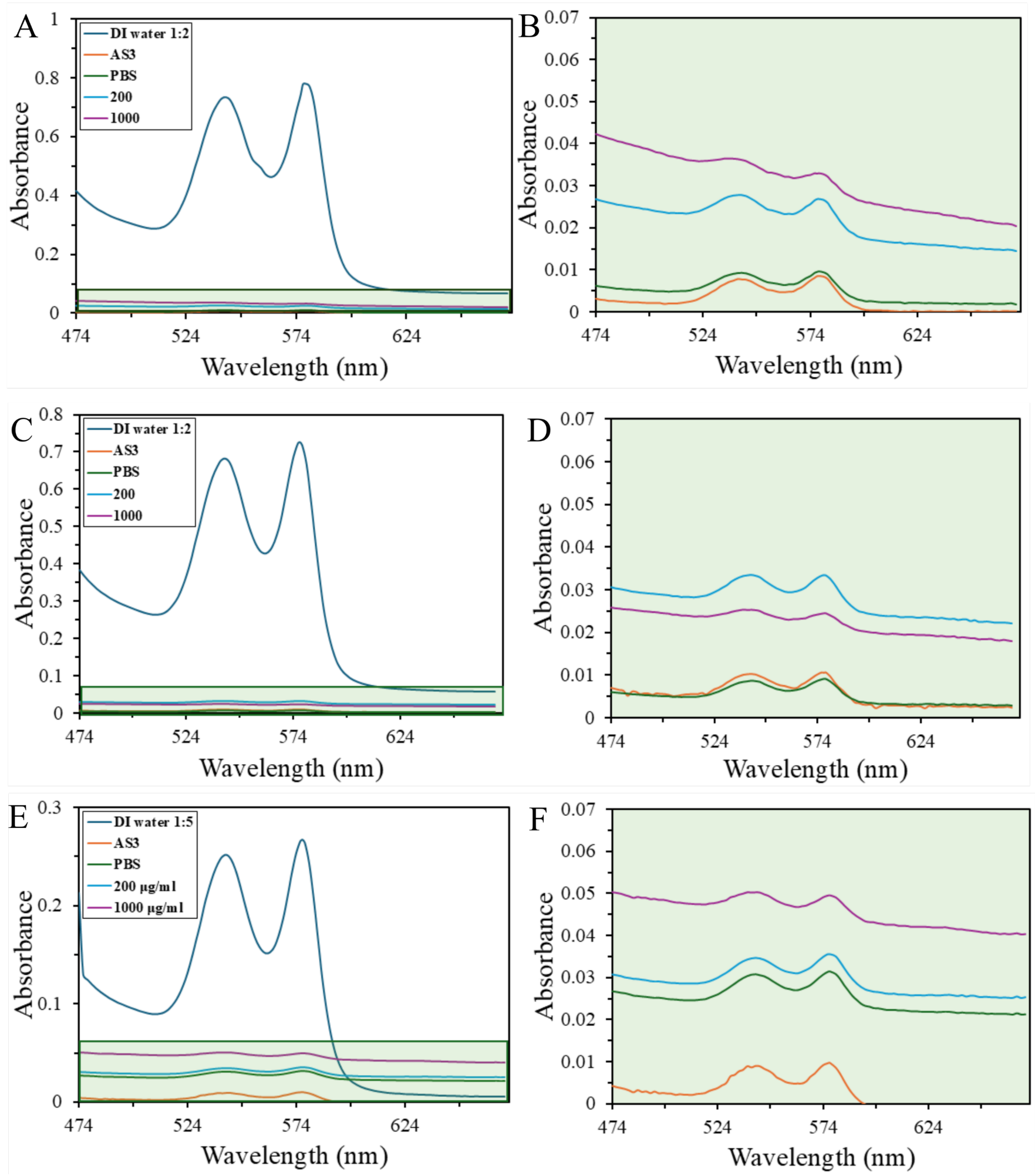

**Figure S8.** UV-Vis absorption spectra for C4 (A), S4 (C), and S3 (E) after 1h incubation. UV-Vis absorption spectra (A, C, E) and magnified hemoglobin region (B, D, F) of supernatants collected after 24 h incubation of RBCs with different samples at 200 and 1000 µg mL$^{-1}$.

**S9. Bright-field microscopy images of SKOV3 cells after 48 h incubation with C35 and C3 nanoparticles at 500 µg mL$^{-1}$**

**Figure S9** shows bright-field optical microscopy images of SKOV3 ovarian cancer cells after 48 h incubation with C35 (top) and C3 (bottom) nanoparticles at 500 µg mL$^{-1}$. Dark deposits are visible in cells incubated with both formulations, consistent with cell-associated nanoparticle accumulation. C35 shows more pronounced and larger dark aggregates relative to C3, consistent with its higher cellular accumulation at the tested concentration.

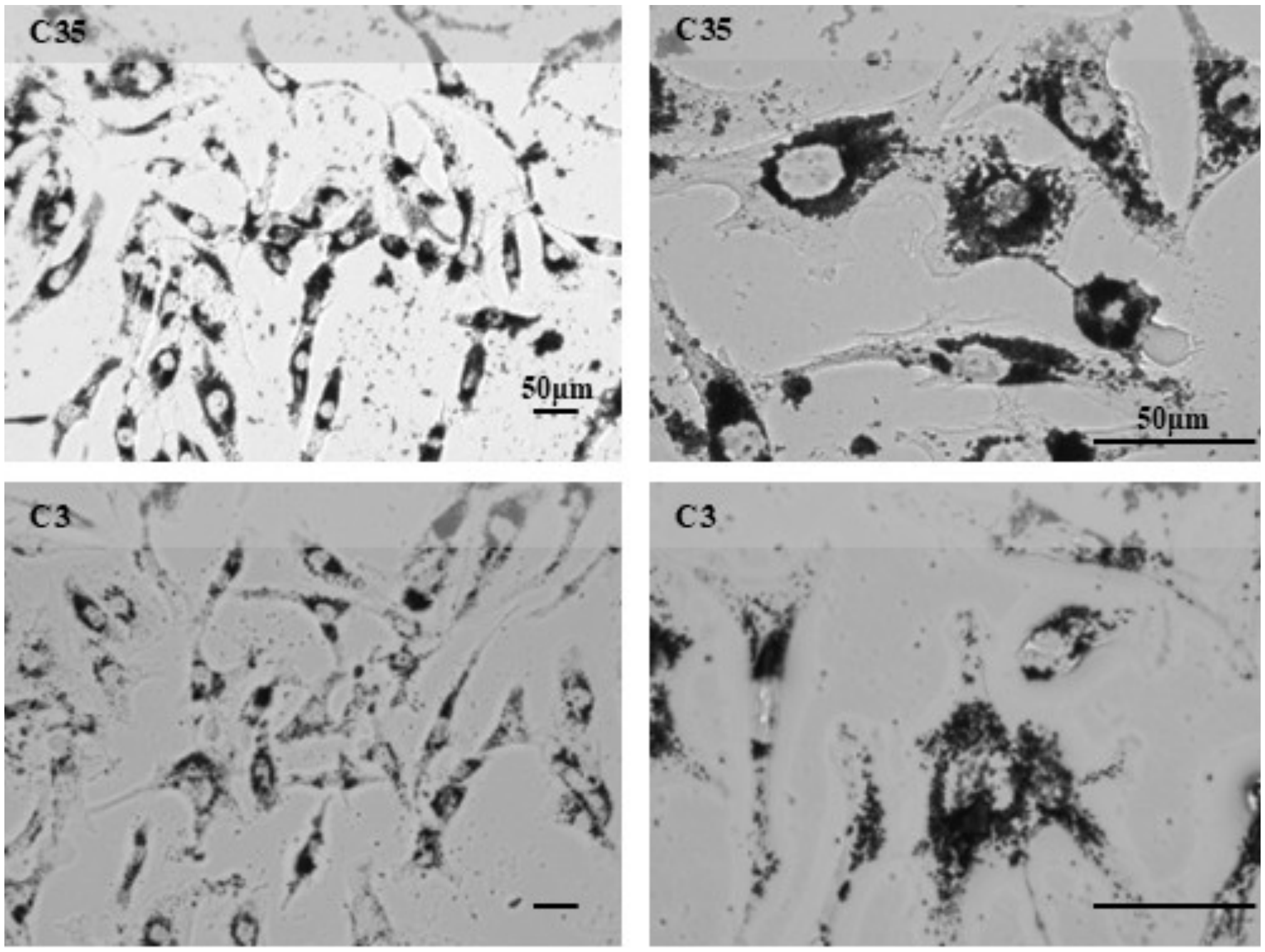

**Figure S9.** Representative bright-field microscopy images of SKOV3 cells after incubation with C35 and C3 ferrite nanoparticles at 500 µg mL$^{-1}$.

**S10. Live/Dead fluorescence images of SKOV3 cells in 3D perfusable GelMA tumor constructs after 48 h perfusion with C35 (A, C) and C3 (B, D) nanoparticles at 300 µg $mL^{-1}$**

**Figure S10** shows Live/Dead fluorescence images of SKOV3 cells in 3D perfusable GelMA constructs after 48 h perfusion with C35 (A, C) and C3 (B, D) at 300 µg $mL^{-1}$. C3 constructs (B, D) show predominantly green fluorescence with minimal red signal, consistent with high cell viability at this concentration. C35 constructs (A, C) display a higher proportion of red cells relative to C3, consistent with the concentration-dependent viability reduction reported for this formulation. These results are consistent with the viability values of 76% and 91% reported for C35 and C3, respectively, at 300 µg $mL^{-1}$.

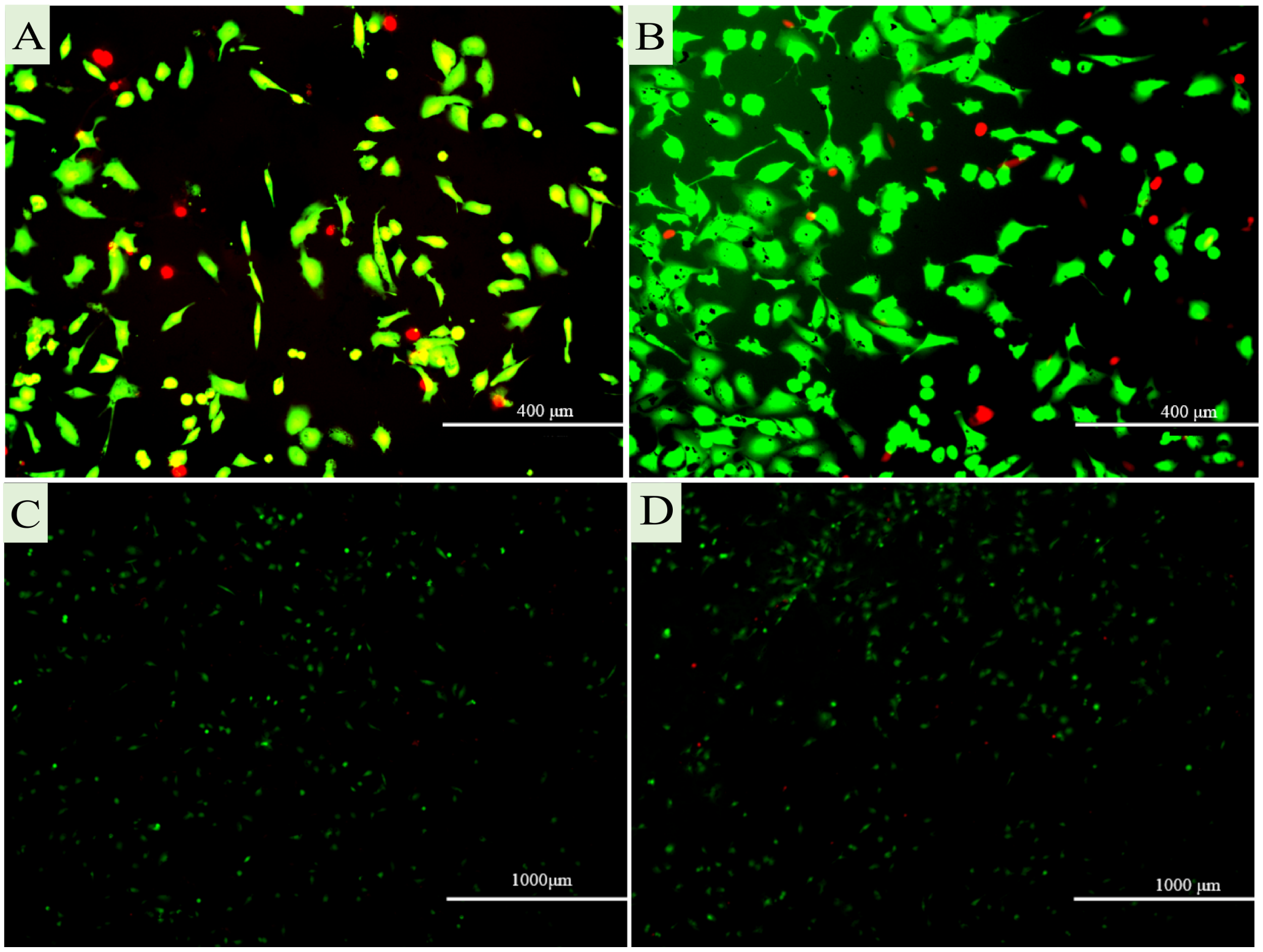

**Figure S10.** Representative Live/Dead fluorescence images of SKOV3 cells in perfused 3D GelMA-gelatin tumor constructs after exposure to (A&D) C35 and (B&D) C3 sample at 300 µg $mL^{-1}$.